\documentclass[prb,twocolumn,amsmath,amssymb,citeautoscript]{revtex4}

\usepackage{amsmath}
\usepackage{amssymb}
\usepackage{epsfig}
\usepackage{graphicx}
\usepackage{wasysym}
\usepackage{bm}

\usepackage{amsmath}
\usepackage{amssymb}
\usepackage{epsfig}
\usepackage{graphicx}
\usepackage{wasysym}
\usepackage{bbm}
\usepackage{subfigure}

\newcommand \be{\begin{equation}}
\newcommand \ee{\end{equation}}
\newcommand \bes{\begin{equation*}} 
\newcommand \ees{\end{equation*}}
\newcommand \bea{\begin{eqnarray}}
\newcommand \eea{\end{eqnarray}}
\newcommand \beas{\begin{eqnarray*}} 
\newcommand \eeas{\end{eqnarray*}}
\newcommand \bfg{\begin{figure}}
\newcommand \efg{\end{figure}}
\newcommand \bfgs{\begin{figure*}} 
\newcommand \efgs{\end{figure*}}
\newcommand \bwt{\begin{widetext}}
\newcommand \ewt{\end{widetext}}

\newcommand \rb{\rm{\bf{r}}}
\newcommand \kb{\rm{\bf{k}}}

\begin{document}

\title{ Band Topology of Insulators via the Entanglement Spectrum }

\author{Ari M. Turner$^{1}$, Yi Zhang$^{1}$, Ashvin Vishwanath$^{1,2}$}

\affiliation{$^1$Department of Physics, University of California,
Berkeley, CA 94720
\\
$^2$Materials Sciences Division, Lawrence Berkeley National
Laboratory, Berkeley, CA 94720 }
\date{Printed \today}

\begin{abstract}
How do we uniquely identify a quantum phase, given its ground state
wave-function? This is a key question for many body theory
especially when we consider phases like topological insulators, that
share the same symmetry but differ at the level of topology. The
entanglement spectrum has been proposed as a ground state property
that captures characteristic edge excitations. Here we study the
entanglement spectrum for topological band insulators. We first show
that insulators with topological surface states will necessarily
also have protected modes in the entanglement spectrum. 
Surprisingly, however, the converse is not true. Protected entanglement modes
can also appear for insulators {\em without} physical surface
states, in which case they capture a more elusive property. This is
illustrated by considering insulators with only inversion symmetry.
Inversion is shown to act in an unusual way, as an antiunitary
operator,  on the entanglement spectrum, leading to this protection.
The entanglement degeneracies indicate a variety of different phases
in inversion symmetric insulators, and these phases
are argued to be robust to the introduction of
interactions.

\end{abstract}

\maketitle

\section{Introduction}
According to the Landau paradigm, phases of matter are classified by
their pattern of symmetry breaking. While this accounts well for a
number of experimentally observed phenomena, several
exceptions have also cropped up. These include the fractional
quantum Hall states and models of gapped spin liquids, which are
characterized by a topological order: for these phases,
the ground state degeneracy
depends on the spatial topology of the sample. Perhaps the simplest
phases for which a topological distinction is present are
topological insulators and superconductors, composed of
noninteracting particles. Experimental manifestations include the
integer quantum Hall effect, spin-orbit induced $Z_2$ topological
insulators in two \cite{Kane,Bernevig,Molenkamp} and three
dimensions \cite{key-2,MooreBalents,Hasan,others}, as well as superfluid
$He^3-B$. Here, although there is a unique ground state and no
symmetry difference from the trivial state, the topological
distinction can manifest itself in different ways. These include a
quantized response function \cite{TKNN} (such as the Hall effect),
or protected surface states. Given the subtleties associated with
identifying topological phases,  a  new tool in this respect would
be welcome.

Recently, it has been shown that studying the entanglement spectrum
is a promising direction to identifying topological phases. Given
the ground state wavefunction, and a partition of the system into a
left and right half, one can perform a Schmidt decomposition:
\begin{equation}
|G\rangle=\sum_{\alpha}\frac{1}{\sqrt{Z}}e^{-\frac{E^e_\alpha}{2}}|\alpha
L\rangle|\alpha R\rangle\label{schmidt}
\end{equation}
Measuring the right half of the system shows it to be in state
$|\alpha R\rangle$ (called a Schmidt state) with probability
$e^{-E^e_\alpha}/Z$. The quantities $E^e_\alpha$ comprise the
entanglement spectrum, and are somewhat like ``energies,"
characterizing how unlikely a given fluctuation is to occur. (Note
that we use the superscript ``e" when describing the entanglement.)
The entropy associated with this probability distribution is the
entanglement entropy, which captures the same features of a phase as
ground 
state
degeneracy (or topological order)\cite{levin}. 
However, the entire spectrum can be
expected to contain more information, allowing one to capture a
wider class of distinctions. Even though entanglement related
quantities are hard to measure directly, they can be conceptually
very useful in identifying phases.

It has been shown for fractional quantum Hall states that although
their bulk is gapped, the entanglement spectrum contains information
about the edge modes. Numerical studies\cite{LiHaldane}
reveal that the largest entanglement eigenvalues (smallest
$E^e_\alpha$) mimic the low-energy spectrum of surface
modes. A relationship of this type is very interesting because the
entanglement spectrum can be calculated entirely from the ground
state wave function. In contrast, the edge modes are excited states
of the Hamiltonian in a sample with boundaries. In a classical
system, one would not be able to look at the ground state and
determine its dynamics. But a quantum mechanical system, even in its
ground state, has some zero-point motion that can give one a sense
of its excited state properties. A direct classification of phases
from their ground state wavefunctions is highly desirable and could
be applied to cases where a candidate ground state (such as a
Gutzwiller projected state) is available.

Here we study the entanglement spectrum in the context of the
simplest class of topological phases - noninteracting topological
band insulators. We show that whenever topological surface states
are present, the entanglement spectrum, as a function of the
momentum parallel to the cut, also has protected entanglement modes.
Essentially, the entanglement spectrum is known to result from
diagonalizing an operator, and this operator can be viewed as a band
insulator Hamiltonian which retains the topology of the physical
insulator. Explicit examples of the entanglement spectrum are worked
out, including an analytical calculation for Landau levels (similar
to the work of Ref.~\onlinecite{rodriguez}) and a
numerical study of three dimensional topological insulators. 
Similar reasoning can be applied to topological superconductors.
Such a relation has also been noticed in isolated examples, like the
quantum Hall \cite{rodriguez} and the $p_x+ip_y$ superconductor
\cite{Noah} edge. Independent studies\cite{Lukasz, HaldaneAPS}
have appeared, pointing to a similar connection.

Next, we consider the converse - i.e. do protected modes in the
entanglement spectrum 
necessarily imply
protected surface excitations? We show explicitly that this is {\em
not} true, by studying an example of a insulator with only inversion
symmetry. Here no protected surface excitations exist, but the
entanglement spectrum features protected states, which points to a
more subtle distinction between phases. The entanglement spectrum
remains gapless because inversion symmetry is retained when dividing
the system for the purposes of calculating entanglement entropy, but
is implemented in a strange way, as a particle-hole symmetry of the
single particle entanglement states. This leads to protected modes.
Thus studying systems with inversion symmetry (or another symmetry
that maps the left and right halves to one another) is a good way to
break the close correspondence between edge states and entanglement,
a point noticed earlier for one-dimensional interacting states in
Ref.~ \onlinecite{frankA,frankB}. When this correspondence breaks
down, the protected entanglement degeneracies still
indicate a distinct phase, although there are no physical surface
states.  Understanding the consequences of this hidden degeneracy
should be interesting.  Although disorder breaks inversion
symmetry, and is hence not normally studied while classifying
topological insulators \cite{shinsei, Kitaev}, clean physical
systems with inversion symmetric bulk states can certainly be
realized.

Finally, we recast the theory given here in a form that is suitable for
studying interacting systems: the action of inversion ($I$) is defined
on the {\em many-body} Schmidt states. It is shown to act as an
anti-unitary operator $IM$, which must satisfy $(IM)^2=+1$ or $(IM)^2=-1$. 
The latter corresponds to the topologically nontrivial
case, and leads to a two fold degeneracy of all states in the
entanglement spectrum. This degeneracy remains on introducing weak
interactions. Part of the topological distinction implied by this
observation is contained in the quantized magnetoelectric
polarizability of insulators, which remains quantized in 
inversion-symmetric systems. We also discuss finer distinctions
between inversion-symmetric insulators, but their physical implications
remain to be identified in future research.




\section{Topological Band Insulators}
A band insulator is described by the single particle Hamiltonian:
\begin{equation}
H = \sum_{\gamma,\rm{\bf{k}}}
\epsilon_\gamma(\rm{\bf{k}})d^\dagger_{\gamma \rm{\bf{k}}}d_{\gamma
\rm{\bf{k}}} \label{bandinsulator}
\end{equation}
where $\rm{\bf{k}}$ is crystal momentum, and $\gamma=1,\dots, N$ is
a band index, of which $n$ bands are filled i.e.
$\epsilon_\gamma(\rm{\bf{k}})<0$ if $1\leq\gamma\leq n$ and
$\epsilon_\gamma(\rm{\bf{k}})>0$ if $n<\gamma \leq N$. The band
wave-functions are:
\begin{equation}
 d^\dagger_{\gamma \rm{\bf{k}}} =
\sum_{\rb}\phi_{\gamma\rm{\bf{k}}}(\rb)\psi^\dagger(\rm{\bf{r}}),
\label{eq:occupied}
\end{equation}
where we have suppressed spin and orbital
indices (These variables can be included
in $\rm{\bf{r}}$ with the spatial coordinates).

Topological phases of band insulators are defined by their \emph{band
topology}.  Two insulators differ topologically
if they cannot be connected by smooth changes of the Hamiltonian
while the band gap remains finite.
It is possible to define
a topological invariant to distinguish
between different phases that depends only on the wave-functions of the
filled bands\cite{TKNN}.

Given a general band insulator,  (\ref{bandinsulator}), a topologically
equivalent insulator can be constructed by setting the energy of all
occupied bands to be equal and negative, and all unoccupied bands to
be equal and positive: e.g.
\begin{eqnarray}
\epsilon_{F\gamma}(\rm{\bf{k}}) &=& -\frac12 \,{\rm if} \,\gamma\leq n\\
\epsilon_{F\gamma}(\rm{\bf{k}}) &=& +\frac12 \, {\rm if}\, \gamma> n.
\end{eqnarray}
Let us call the corresponding operator $\hat{Q}$. Note, this is
related to the projection operator $\hat{P}$ on the filled bands via
$\hat{Q}= \frac12{\bf 1} - \hat{P}$.
This ``flat band" limit has been found to be useful in the
classification of topological insulators \cite{shinsei}, and will be
used here as well. An important connection is with the correlation
function $C(\bf{r},\bf{r'})=\langle \psi^\dagger_{\bf{r}'} \psi_{\bf{r}}
\rangle$, evaluated
in the ground state. Now $\hat{P}=\hat{C}$, i.e.
the correlation function of the band insulator, viewed as a matrix,
is simply the projection operator onto the filled bands.
Hence
\begin{equation}
\hat{Q}=\sum_{\bf{r}}\frac 12 \psi_{\bf{r}}^\dagger\psi_{\bf{r}}-\sum C(\bf{r},\bf{r}')\psi_{\bf{r}}^\dagger\psi_{\bf{r}'}.
\label{Q}
\end{equation}
describes the flat-band insulator.
Since
correlations in a gapped state like a band insulator fall off
exponentially rapidly with separation, the effective ``flat band"
Hamiltonian (\ref{Q}) has essentially short ranged matrix elements,
as for a physical operator. Hence, it can be viewed also as a
bona fide Hamiltonian, with the same band topology as the starting
Hamiltonian in Eqn. \ref{bandinsulator}.

Most band insulators with nontrivial band topology are characterized
by edge states. Consider representing the Hamiltonian of a band
insulator in real space: \bea H &=&
\sum_{\rb\rb'}\mathcal{H}(\rb,\rb')\psi^\dagger(\rb)\psi(\rb') \\
\mathcal{H}(\rb,\rb') &=&
\sum_{\gamma\kb}\epsilon_\gamma(\kb)\phi^*_{\gamma\rm{\bf{k}}}(\rb)\phi_{\gamma\rm{\bf{k}}}(\rb')
 \label{bandwavefns}\eea
A boundary along the plane $x=0$, with the physical system to the
right ($x>0$), is obtained by truncating the Hamiltonian: \bea
\mathcal{H}^R(\rb,\rb')&=&0 \, \rm{if} \, x<0 {\, \rm or} \, x'<0\\
\mathcal{H}^R(\rb,\rb')&=& \mathcal{H}(\rb,\rb') \, {\rm
otherwise}\eea The spectrum of states localized deep in the bulk is
unaffected by the cut, due to the presence of a finite gap. However,
as one approaches the edge, it is possible that states appear within
the bulk gap. Most topologically nontrivial insulators have
``protected" edge states throughout the bulk gap;
they cannot be removed by smooth changes of the Hamiltonian.
Thus, if the
Hamiltonian (\ref{bandinsulator}) has protected edge modes, so
must its ``flat band" version (\ref{Q}), when restricted to a half
space. 

\section{The Entanglement Spectrum}
\begin{figure}
\includegraphics[width=0.35\textwidth]{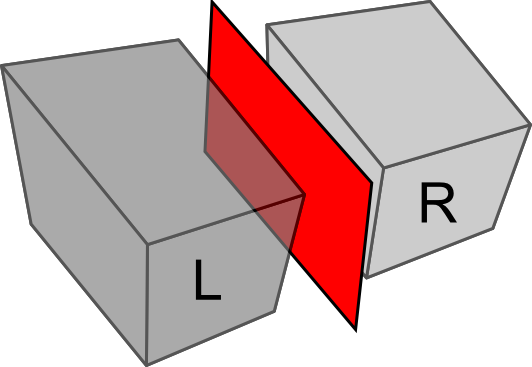}
\caption{A cut that divides the physical system into two halves.}
\label{fig:cut}
\end{figure}

Now we will review what the entanglement spectrum is, and show that
the edge modes of the flat band Hamiltonian are in fact edge modes of
the entanglement spectrum.

Consider the ground state wavefunction of a general Hamiltonian
$H$, $\Psi_G(\{a^R_i\},\,\{a^L_i\})$, where $\{a^R_i\}$
($\{a^L_i\}$) specifies the configuration of the system to the right
(left) of an imaginary cut. Tracing out the degrees of freedom on
the left, one obtains the density matrix:
\begin{equation}
\rho_R(\{a^R_i\},\,\{\tilde{a}^R_i\}) = \sum_{\{a^L_i\}} \Psi_G(\{a^R_i\},\,\{a^L_i\})\Psi_G^*(\{\tilde{a}^R_i\},\,\{a^L_i\}).
\end{equation}
For a normalized wavefunction, ${\rm Tr}[{\bf \rho}_R]=1$, and the eigenvalues of ${\bf \rho}_R$ lie in the range $[0,\,1]$. Thus we can write:
\begin{equation}
{\bf \rho}_R= Z^{-1} e^{-H^e_R}; \,\, Z={\rm Tr} e^{-H^e_R}.
\label{boltzmann}
\end{equation}
We will call $H^e_R$ the entanglement ``Hamiltonian". 
The eigenvalues of the density matrix are denoted by $\rho_\alpha = Z^{-1}e^{-E^e_\alpha}$, where the set $E^e_\alpha$ is termed the ``entanglement spectrum". Note, these ``energies" have nothing to do with the physical energies of the system - they refer entirely to properties of the ground state. (Note that $Z$ is introduced
in order to allow us to set the zero of ``energy" conveniently.)

If both the system and the cut preserve translation symmetry, one
can simultaneously diagonalize the density matrix and the
translation operators parallel to the cut $T_x, \, T_y$. The
entanglement spectrum is then obtained as a function of momentum
$\mathbf{K}_\perp$ along the cut: $E^e_\alpha(\mathbf{K}_\perp)$.
This object can
capture fairly detailed properties of the ground state wavefunction
that are discussed in what follows.

For the case of {\em single particle} Hamiltonians,
such as Eqn. \ref{bandinsulator}, Ingo Peschel has shown
how to determine the entanglement Hamiltonian and its eigenvalues\cite{peschel}.
The first step is to realize that
the entanglement Hamiltonian must be quadratic in the Fermionic
operators:
\begin{equation}
H^e_R = \sum_{r,r'} \psi^\dagger_r
{{\verb"H"}^R_{rr'}}\psi_{r'}. \label{singleparticleH}
\end{equation}
This can be seen, and the single particle ``entanglement
Hamiltonian" ${\verb"H"}^R$ determined, by noting that
any
correlation function of operators that only involve degrees of
freedom on the right is unaffected by tracing out the left half of
the system.
For example,
consider the
correlation function $C(\bf{r},\bf{r}')=\langle
\psi^\dagger_{\bf{r}'}\psi_{\bf{r}}\rangle$, in the ground state, before the
system is cut. Now, if both coordinates $\bf{r},\,\bf{r}'$ belong to the right
half of the system, then they are unaffected on tracing over the
left half. Therefore, the same result should be obtained by using
the density matrix for the right half,
\begin{equation}
{\rm Tr} [{\bf \rho^R}\psi^\dagger_{\bf{r}'}\psi_{\bf{r}}]=C_R(\bf{r},\bf{r}')
\label{Correlation}
\end{equation}
where $C_R$ is the correlation function restricted to this half,
namely $C_R(\mathbf{r},\mathbf{r}')=C(\mathbf{r},\mathbf{r}')$
if both
coordinates are on the right, and zero otherwise. A similar
argument applies to multiparticle correlations.  Thus,
since Wick's theorem is obeyed by
correlation functions in the original uncut system, it will continue to
hold for just the right half of the system in the mixed state
$\bf \rho^R$.
Consequently, $\bf \rho^R$ must be Gaussian, i.e. the exponential of a
quadratic form of Fermion operators. Requiring also conservation of
particle number leads to the general form in Eqn.
\ref{singleparticleH}.
An expression for ${\verb"H"}^R$ may be obtained by returning to two-point
correlations, and requiring
that the ``Boltzmann" distribution, Eq. (\ref{boltzmann}), gives
the expression Eq. (\ref{Correlation}).
One
expands the Fermion operators in a basis which diagonalizes
${\verb"H"}^R$; in this basis, $C$ becomes a diagonal matrix recording
the mean 
occupation numbers of the states,
given by the Fermi distribution. Thus,
$\hat{C}_R = [1+e^{{\verb"H"}^R}]^{-1}$, or
\begin{equation}
{\verb"H"}^R = \log \left [\frac{{\bf
1}-\hat{C}_R}{\hat{C}_R}\right ]. \label{singleparticleHEqn}
\end{equation}
Equivalently, the operator
$\hat{Q}_R= \frac12 {\bf 1} - \hat{C}_R$,
satisfies the relation:
$\hat{Q}_R= \frac12 \tanh [\frac12
{{\verb"H"}^R}]$. Thus if $p_i$ are the eigenvalues of
$\hat{C}_R$, which are readily seen to satisfy $0\leq p_i \leq
1$, then the eigenvalues $\epsilon^e_i$ of the `single particle'
entanglement Hamiltonian ${\verb"H"}^R$ are obtained from
Eqn.\ref{singleparticleHEqn}:
\begin{equation}
\epsilon^e_i = \log \left [\frac{1-p_i}{p_i} \right ]
\label{eigenvalues}
\end{equation}

The many body entanglement spectrum $E^e_\alpha$ is obtained by
deciding the occupancy of the single Fermion modes, so that mode $i$
has $n^\alpha_i=0,\,1$. Then $E^e_\alpha = \sum_in^\alpha_i\epsilon^e_i$.
The density matrix eigenstate with the largest eigenvalue
is like the Fermi sea of a physical
system; it corresponds to the lowest
``energy" $E^e_\alpha$ obtained by filling all $\epsilon^e_i<0$.

For a translationally symmetric cut the entanglement energies
can be resolved as a function of the total transverse momentum $\bf{K}_\perp$
(following Ref.~ 
\onlinecite{LiHaldane}). These energies can again be built from the
single particle ``energies" of the occupied states,
$\epsilon^e_i(\bf{k}_\perp)$, which are functions of the momentum
parallel to the cut. The total momentum $\bf{K}_\perp$ is
also just the sum of the momenta of the filled single particle
states. Thus, the fundamental object we will focus on calculating is
the single particle ``spectrum" as a function of transverse momentum
$\epsilon^e_i(\bf{k}_\perp)$.

\section{Connecting the Entanglement Spectrum to Edge Modes}
The previous discussion clarifies the precise connection between
the entanglement spectrum and the edge spectrum of an insulator.
The single particle entanglement spectrum
$\epsilon^e_i(\bf{k}_\perp)$ is related to the spectrum obtained when a flat
band version of a physical Hamiltonian is studied in the presence
of an edge. The eigenvalues of the flat band Hamiltonian $\hat{Q}$,
which are $\frac12-p_i(\mathbf{k}_\perp)$, are related to the
entanglement spectrum $\epsilon^e_i(\mathbf{k}_\perp)$ via Eqn.
(\ref{eigenvalues}), or \begin{equation}\frac12-p_i(\mathbf{k}_\perp)=
\frac12 \tanh [\frac12
{\epsilon^e_i(\mathbf{k}_\perp)}].\label{eigenvaluesQ}\end{equation}
Most states have eigenvalues $\epsilon_i\rightarrow\pm\infty$, because
of the nonlinear relationship;
the eigenvalues of
$\hat{C}$ in the bulk are $1$ and $0$.
Very negative eigenvalues represent bulk states that are
occupied, and large positive eigenvalues represent unoccupied states
\footnote{In fact,
every entanglement eigenstate with a finite $\epsilon^e$ is a bound
state of the surface, since the bulk dispersions are flat and equal
to $\pm\infty$.  States extend farther into the bulk as $|\epsilon^e|$
increases.}.

\emph{In cases where topologically protected surface
states of the physical Hamiltonian are expected,
the flat band deformation which is
topologically equivalent is also expected to have surface states},
filling the whole gap between the bulk states at $+\infty$ and $-\infty$.

Note however, the converse is not necessarily true.
For inversion symmetry (see below), the edge
of the flat band Hamiltonian has structure which
signifies protected features of a phase, although this structure
is not present in the physical surface states of
a generic Hamiltonian in the same topological class.

We will now illustrate our reasoning with various examples.

\section{Examples}
\label{sec:Examples} Let us summarize the procedure for obtaining the
entanglement spectrum. The following prescription takes advantage of
${\rm \bf k}_\perp$ conservation to reduce the problem to a one-dimensional
problem:
 (i) obtain the correlation function
restricted to the right half, by summing over occupied bands. If the
wavefunction of the filled orbitals at momentum ${\rm \bf
k}=(k_x,\,\bf{k}_\perp)$ is $\phi_{i{\rm \bf k}}e^{i{\rm \bf k}\cdot{\rm
\bf r}}$, where ${\rm \bf r}$ is the unit cell position, and $a$
refers to other indices such as sublattice and spin, then:
$$C^R_{\bf{k}_\perp}(x,a;x',a') =
\sum_{i,\,k_x}e^{ik_x(x-x')}\phi^*_{i{\rm \bf k}}(a')\phi_{i{\rm \bf
k}}(a).$$(ii) Find $C$'s eigenvalues $p_i(\bf{k}_\perp),$ by
solving the eigenvalue equation
$$\sum_{x'a'}C^R_{\bf{k}_\perp}(x,\,a;x',\,a')f^R_{i{\bf k}_\perp}(x',\,a')=p^i({\bf k}_\perp)f^R_{i{\bf k}_\perp}(x,\,a)$$
(iii) Then the single particle entanglement spectrum
$\epsilon^e_i(\bf{k}_\perp)$ can be read off from equation
(\ref{eigenvaluesQ}) and the eigenfunctions are $f_{i\mathbf{k}_\perp}(x)e^{i\mathbf{k}_\perp\cdot\mathbf{r}_\perp}$.
We will graph our results for
$\frac12-p^i(\bf{k}_\perp)$ instead of $\epsilon^e_i(\mathbf{k}_\perp)$,
since they are monotonically related.

\subsection{Integer Quantum Hall:}

\begin{figure}
\includegraphics[width=0.35\textwidth]{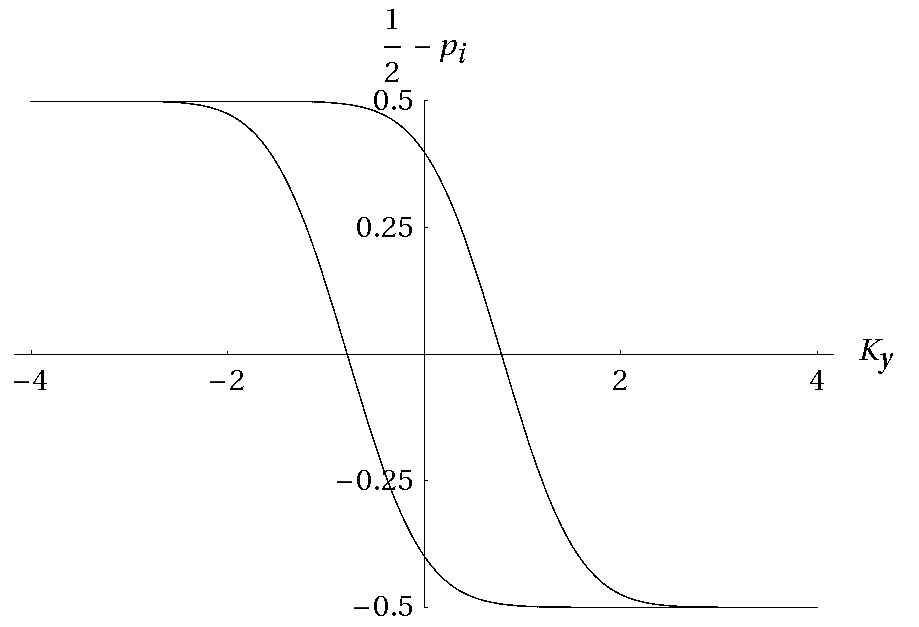}
\caption{Entanglement edge states for the $\nu=2$ integer quantum Hall state.}
\label{fig:IQHE}
\end{figure}

The first example is based on the
work of Ref.~ \onlinecite{rodriguez} on the integer quantum Hall effect.
This is an especially clear illustration of the general relationship between
the correlation function and the entanglement Hamiltonian.
Consider electrons confined to a plane and in a strong magnetic
field; assume there is no crystal potential and suppose
that exactly $\nu$ Landau levels are fully occupied. We
show that the entanglement spectrum contains the expected $\nu$
chiral edge modes. Using the Landau gauge, $A_y=Bx$, and labeling
the states by the momentum along the cut $k_\perp=k_y$, the
restricted correlation function is given by: $C^R_{k_y}(x,\,x') =
\sum_{n=1}^\nu \phi^*_{nk_y}(x')\phi_{nk_y}(x)$ where the normalized
eigenstates are $\phi_{nk_y}(x)={\rm H}_n(X-K_y)e^{-(X-K_y)^2/2}$;
here $X=x/l;\,K_y=k_yl$ are dimensionless variables scaled by the
magnetic length $l=\sqrt{\hbar/eB}$, and ${\rm H}_n$ are
appropriately normalized Hermite polynomials.

Now, an eigenfunction of $C^R_{k_y}$ must be constructed from linear
combinations of the Landau level states $f^a = \sum_{m=1}^\nu c_m^a
\phi_m$ (where $k_y$ has been suppressed). The coefficients
$c_m^a$ are easily seen to be eigenvectors of the $\nu\times\nu$
matrix $F_{nm}(k_y)=\int_0^\infty
dx\,\phi^*_{nk_y}(x)\phi_{mk_y}(x)$. For $\nu=1$, this is a number,
which is the eigenvalue itself. Thus
$$1/2-p_0(k_y)=-\frac12{\rm erf}(K_y),$$ which interpolates
between $-1/2$ (when $k_y=-\infty$) and $+1/2$ (when $k_y=+\infty$).
This is the single chiral mode, as expected. Similarly, the $\nu=2$
case can be solved analytically, the result for the two eigenvalues
is now
$$\frac12-p_{1,\,2}=-\frac12[{\rm erf}(K_y)-e^{-K_y^2}\left(\frac{K_y}{2\sqrt{\pi}}\pm \sqrt{\frac{2+K_y^2}{\pi}}\right)]$$
leading to two modes as shown in the figure.
These two modes are like the dispersion of two
chiral particles in a one-dimensional system.
In general, the entanglement Hamiltonian of a gapped system
behaves like the Hamiltonian
of a system with one dimension fewer.
There may be infinitely many bands besides those near
$\epsilon^e=0$, but their energies
rapidly approach $\pm\infty$.

\subsection{Three Dimensional Topological Insulator with Time Reversal
Symmetry:} We now calculate the entanglement spectrum for a three
dimensional topological band insulator, an insulator with surface
modes protected by time reversal symmetry.
By explicit calculation we confirm the expectation that these
modes are also captured by the entanglement spectrum, both in the
case of strong and weak topological insulators.

We consider the model of a topological insulator on the diamond
lattice introduced in Ref.~\onlinecite{key-2}, given by the Hamiltonian:
\begin{eqnarray}
H =  \sum_{ ij} t_{ij}c^\dagger_{i\sigma} c_{j\sigma}
+8it_{SO}\sum_{\langle\langle
ik\rangle\rangle}c^\dagger_{i\sigma}(\mathbf{d}^{\,1}_{ik}\times
\mathbf{d}^{\,2}_{ik})\cdot \bm{\sigma}_{\sigma\sigma'} c_{k\sigma'}
\label{Eq:FuKaneDiamond}
\end{eqnarray}
where the first term contains the four nearest neighbor hopping
elements which are taken as $t$ for three of the bonds, and
$t+\delta t$ for the fourth bond oriented along the (-1,1,-1)
direction. To access a generic Hamiltonian we also include regular
second-neighbor hopping $t_2$ (not shown in Eq. \ref{Eq:FuKaneDiamond})
in all directions. The spin orbit interaction $t_{SO}$
appears in the second
term, inducing hopping between second neighbor sites. In this term,
$\mathbf{d}^{\,1}_{ik},\, \mathbf{d}^{\,2}_{ik}$ are the two nearest
neighbor bond vectors leading from site $i$ to $k$, and
$\bm{\sigma}$ are the spin Pauli matrices. 
This model respects time
reversal symmetry, and also inversion symmetry (i.e.
$\mathbf{r}-\mathbf{r}_0\rightarrow \mathbf{r}_0-\mathbf{r}$, about
a bond center $\mathbf{r}_0$). Without the second-neighbor hopping,
the model's bulk and surface spectra display the non-generic
feature that all energy levels come in $\pm E$ pairs at each
momentum.

\begin{figure}
\includegraphics[width=0.20\textwidth]{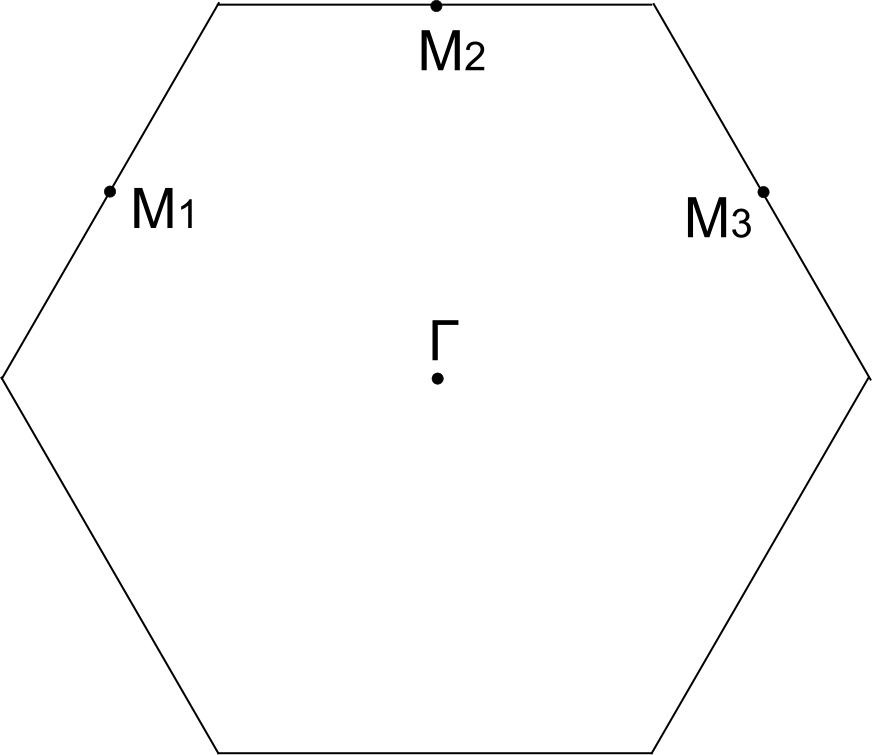}
\caption{Brillouin zone of the (1,1,1) surface of the diamond
lattice. The labelled points are time reversal invariant momenta.}
\label{fig:BZ}
\end{figure}

\begin{figure}[h!]
\begin{center}
\subfigure (a) 
{
    \label{fig:STIdispersion}
    \includegraphics[width=0.35\textwidth]{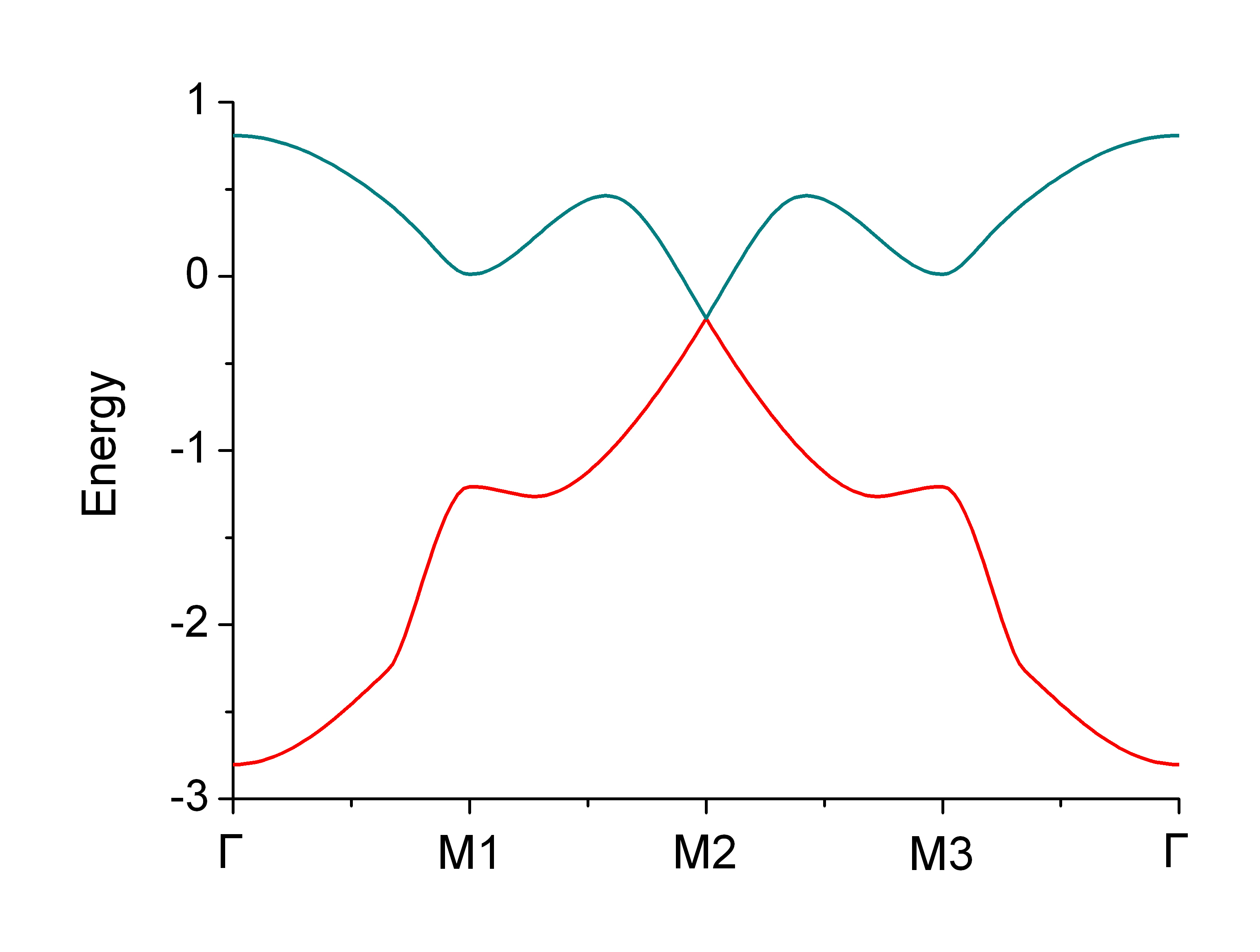}
} \hspace{2cm}
\subfigure (b) 
{
    \label{fig:STIspectrum}
    \includegraphics[width=0.35\textwidth]{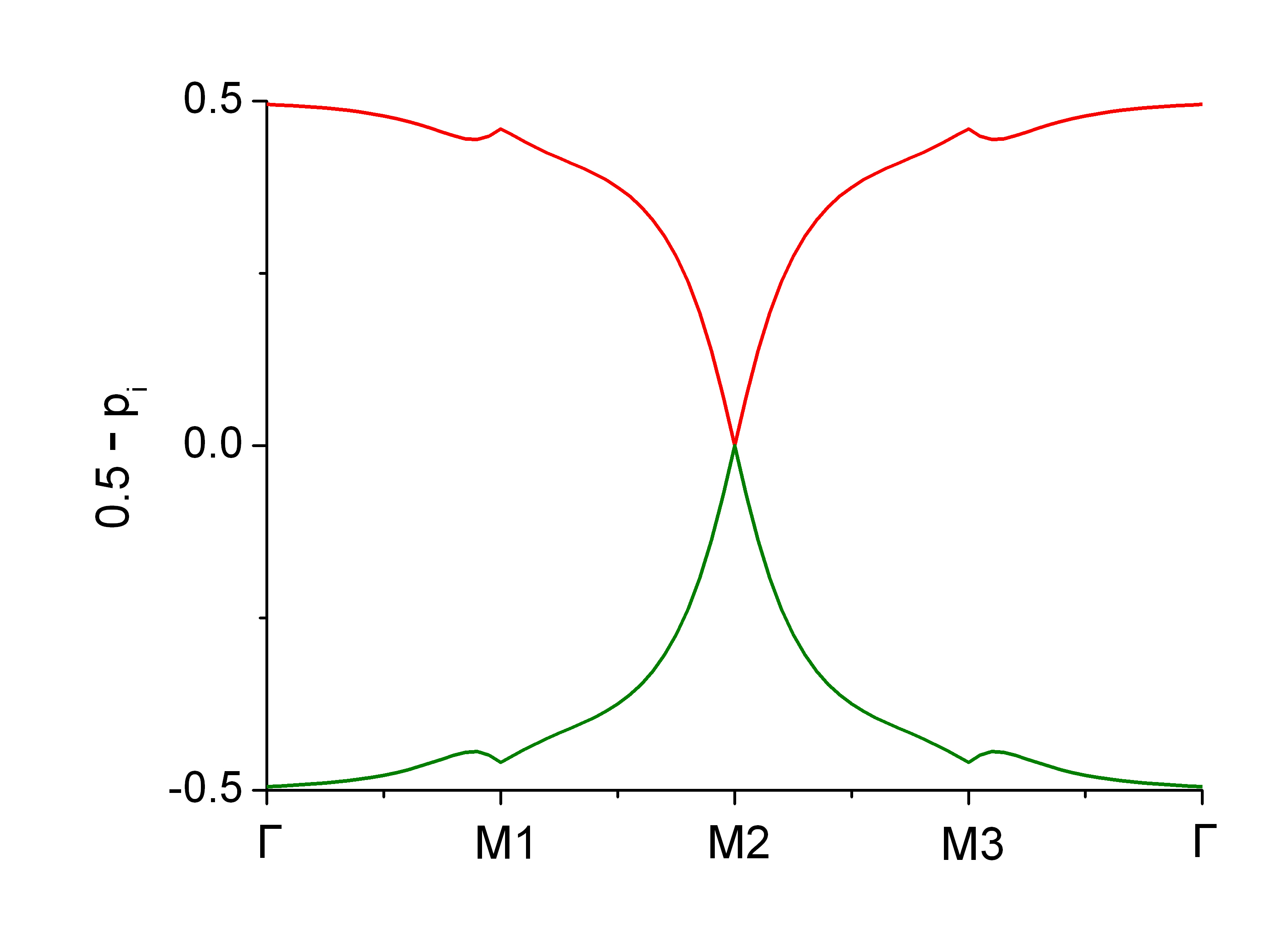}
} \caption{Strong topological insulator. a) The dispersion
of the physical surface states. Note that the node
is not at zero energy, since particle-hole symmetry is broken.
b) The entanglement spectrum, $\frac{1}{2}-p$, which has the
node at zero $\epsilon^e$, in spite of particle-hole symmetry being
broken.}
\label{fig:STI} 
\end{center}
\end{figure}
We now compare the physical surface states that appear in this
model and the entanglement spectrum of a cut with the same
orientation. We choose a surface normal to the (1,1,1) direction so
as to cut bonds with strength $t$. One obtains a strong topological
insulator, with an odd number of surface Dirac nodes if $\delta t>0$
(assuming $t>0$). The surface states computed for a single surface
are shown in Figure \ref{fig:STI}(a), and a single Dirac node
characteristic of the strong topological insulator is obtained. The
node is centered at the time reversal invariant $M2$ point of the
surface Brillouin Zone (BZ) (see Figure \ref{fig:BZ}). This is
selected by the direction of the strong bond $t+\delta t$. Note
that the surface spectrum is not symmetric between positive and
negative energies, as it would be if the model
were particle-hole symmetric.

Below, in Figure \ref{fig:STI}(b), we display the two surface
eigenvalues of the
single particle entanglement spectrum $\epsilon^e_i(\bf{k}_\perp)$,
obtained from the ground state of the Hamiltonian in Eqn.
\ref{Eq:FuKaneDiamond} by dividing the system into two halves. The
plane that divides the system is oriented in the same way as the
physical boundary previously discussed. It is more convenient to
display $\frac12\tanh(\frac{\epsilon^e_i(\bf{k}_\perp)}{2})$ which is
related to the eigenvalues $p_i(\bf{k}_\perp)$ of the correlation
function $C_R$ via $\frac12\tanh(\frac{\epsilon^e_i(\bf{k}_\perp)}{2})=
\frac12 -p_i(\bf{k}_\perp)$. Note zero ``energy",
$\epsilon^e_i(\mathbf{k}_\perp)=0$, also corresponds to the
zero of $\frac{1}{2}-p_i$. Clearly, we see that the entanglement spectrum also
displays the same characteristics as the protected surface modes.
There is a single Dirac cone which is centered at the
same point in the surface BZ as the surface state itself. The ``energy"
of this Dirac cone is curiously zero (understanding this is our next job).

\begin{figure}[h!]
\begin{center}
\subfigure (a) 
{
    \label{fig:WTIdispersion}
    \includegraphics[width=0.35\textwidth]{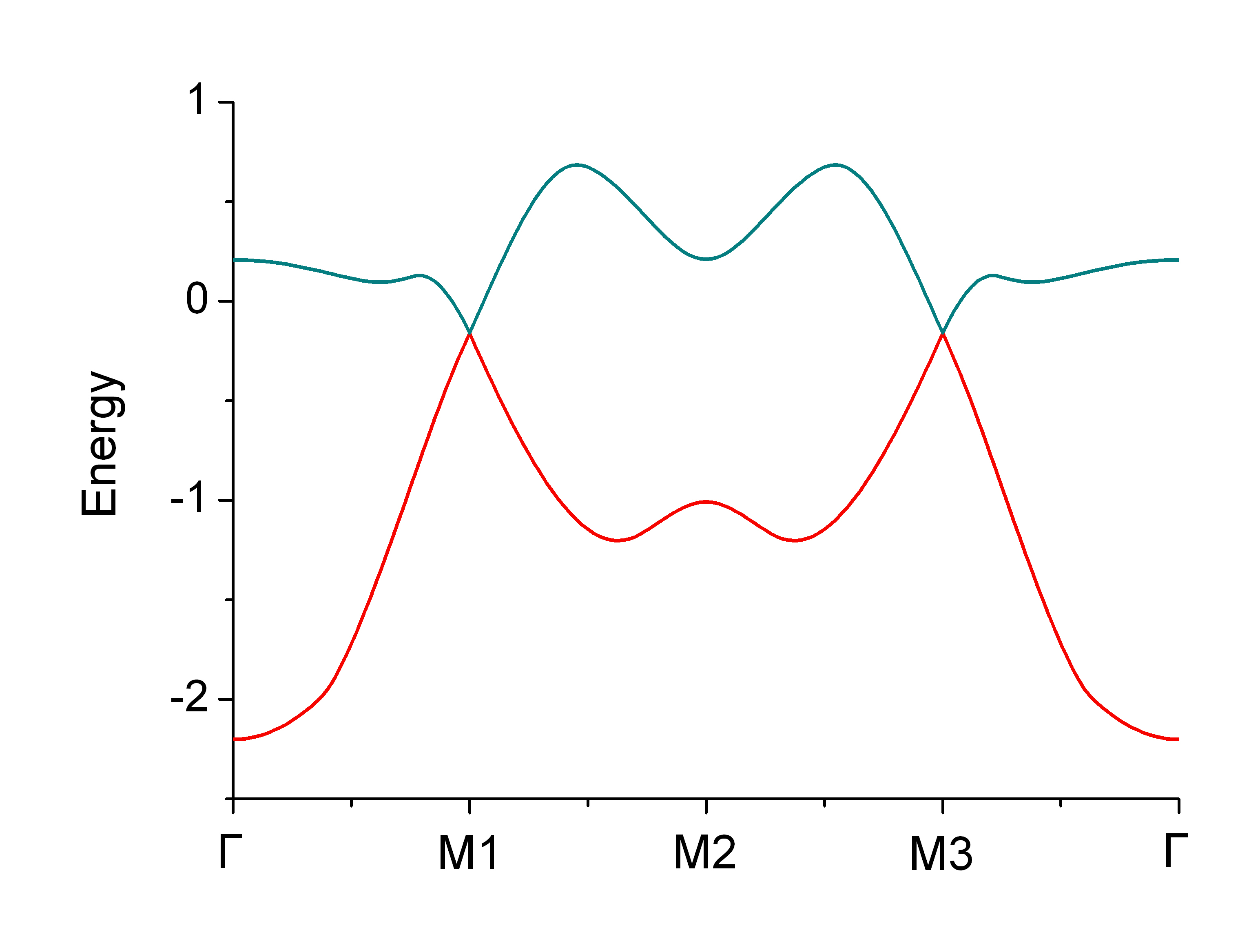}
} \hspace{2cm}
\subfigure (b) 
{
    \label{fig:WTIspectrum}
    \includegraphics[width=0.35\textwidth]{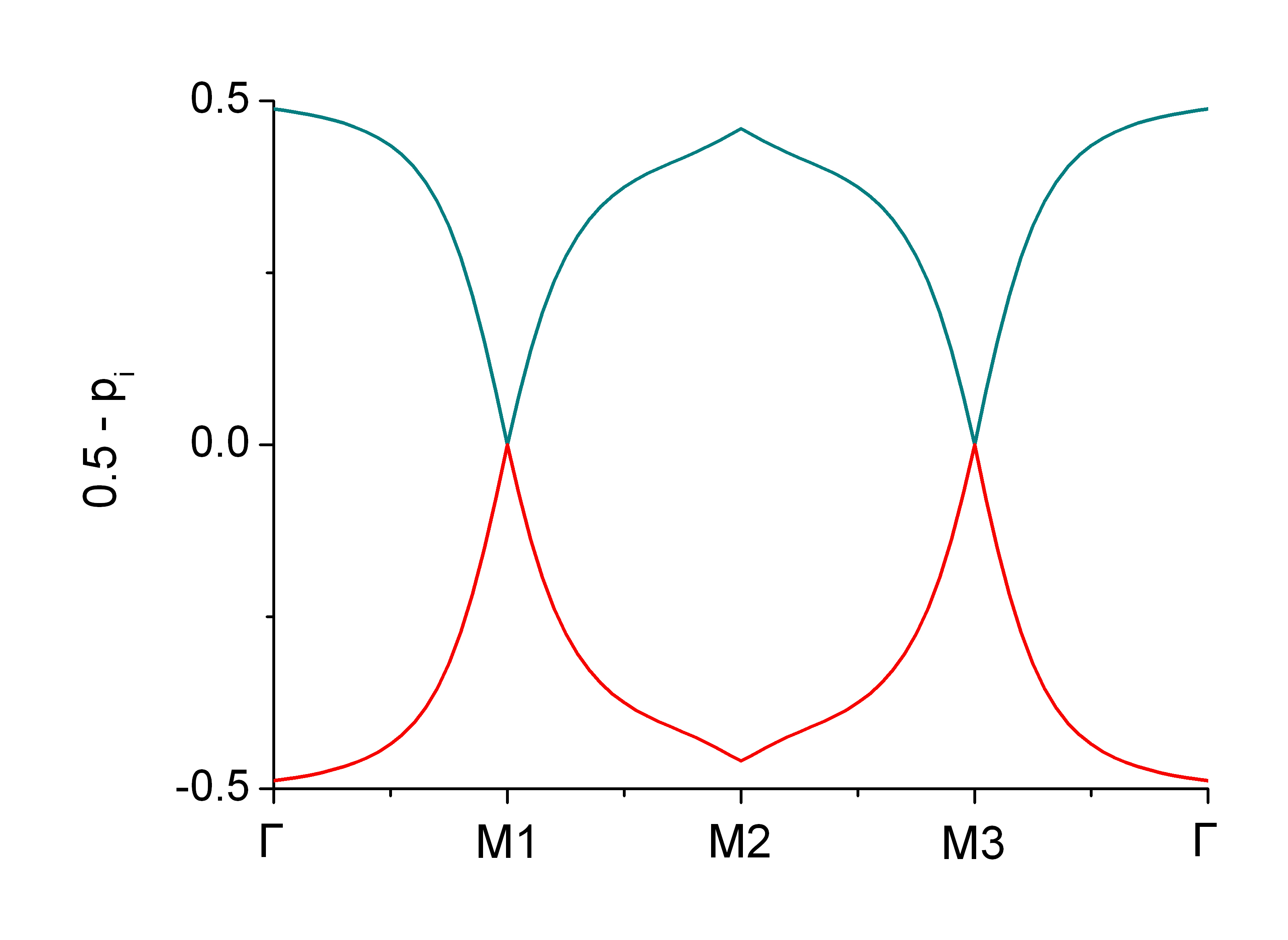}
} \caption{Weak topological insulator. a) The dispersion
of the physical surface states. b) The entanglement spectrum.}
\label{fig:WTI} 
\end{center}
\end{figure}

Similarly, if we choose $\delta t<0$, we obtain a weak topological
insulator, whose physical surface states and entanglement spectrum
are compared in Figure \ref{fig:WTI}, for the same surface as above.
Again, the Dirac nodes of the physical surface states, and their
location in the Brillouin zone, which are fixed by band topology,
are captured by the entanglement spectrum.

We now briefly describe the computation that gives us the
entanglement spectrum in the figures. Essentially, we follow the three
step procedure outlined in Section \ref{sec:Examples}, where
$\bf{k}_\perp$ is a momentum on the surface BZ. Instead of dealing with a
pair of semi-infinite systems $R,\,L$, it is more convenient to
divide the system into three parts, a central piece, whose reduced
density matrix is computed, and one part to the right and one to the
left, which one traces over. This gives us two entanglement
surfaces, but since they are well separated by a gapped region, the
spectra are essentially independent, and we only plot those whose
wavefunctions are concentrated near one of the cuts. For the
numerics we have taken the separation to be 80 unit cells thick, and
we have taken the following
parameters in the Hamiltonian Eqn. (\ref{Eq:FuKaneDiamond}): $t=1.0$,
$t_{SO}=0.125$, $\delta t=1$ $(-0.9)$, $t_2=0.1$ for the strong (weak)
TI. The bulk energy bands as well as the entanglement spectrum
eigenvalues near $1/2-p=\pm 1/2$ are not shown for clarity.

\begin{figure}[h!]
\begin{center}
\subfigure (a) 
{
    \label{fig:noTdispersion}
    \includegraphics[width=0.35\textwidth]{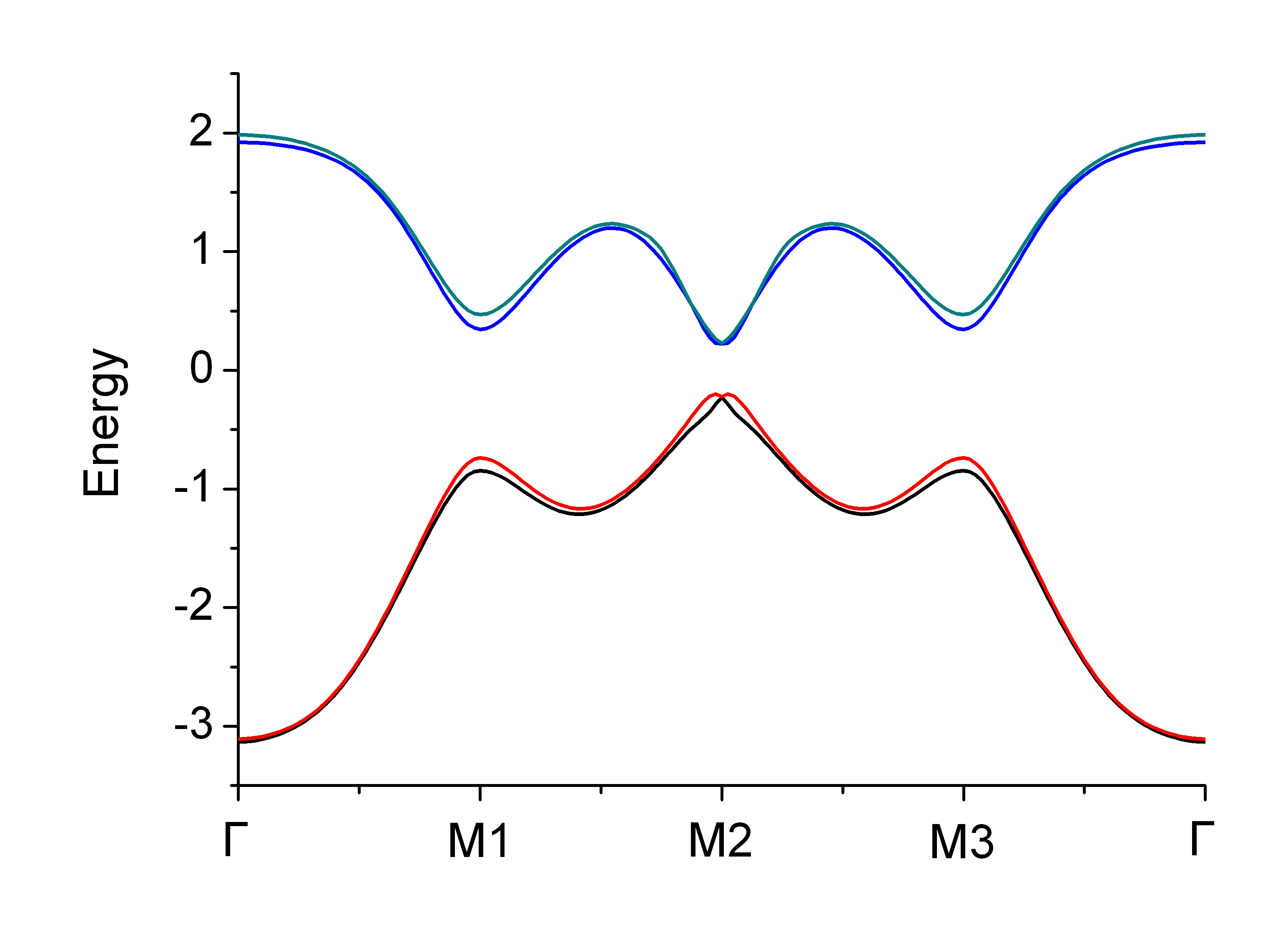}
} \hspace{2cm}
\subfigure (b) 
{
    \label{fig:noTspectrum}
    \includegraphics[width=0.35\textwidth]{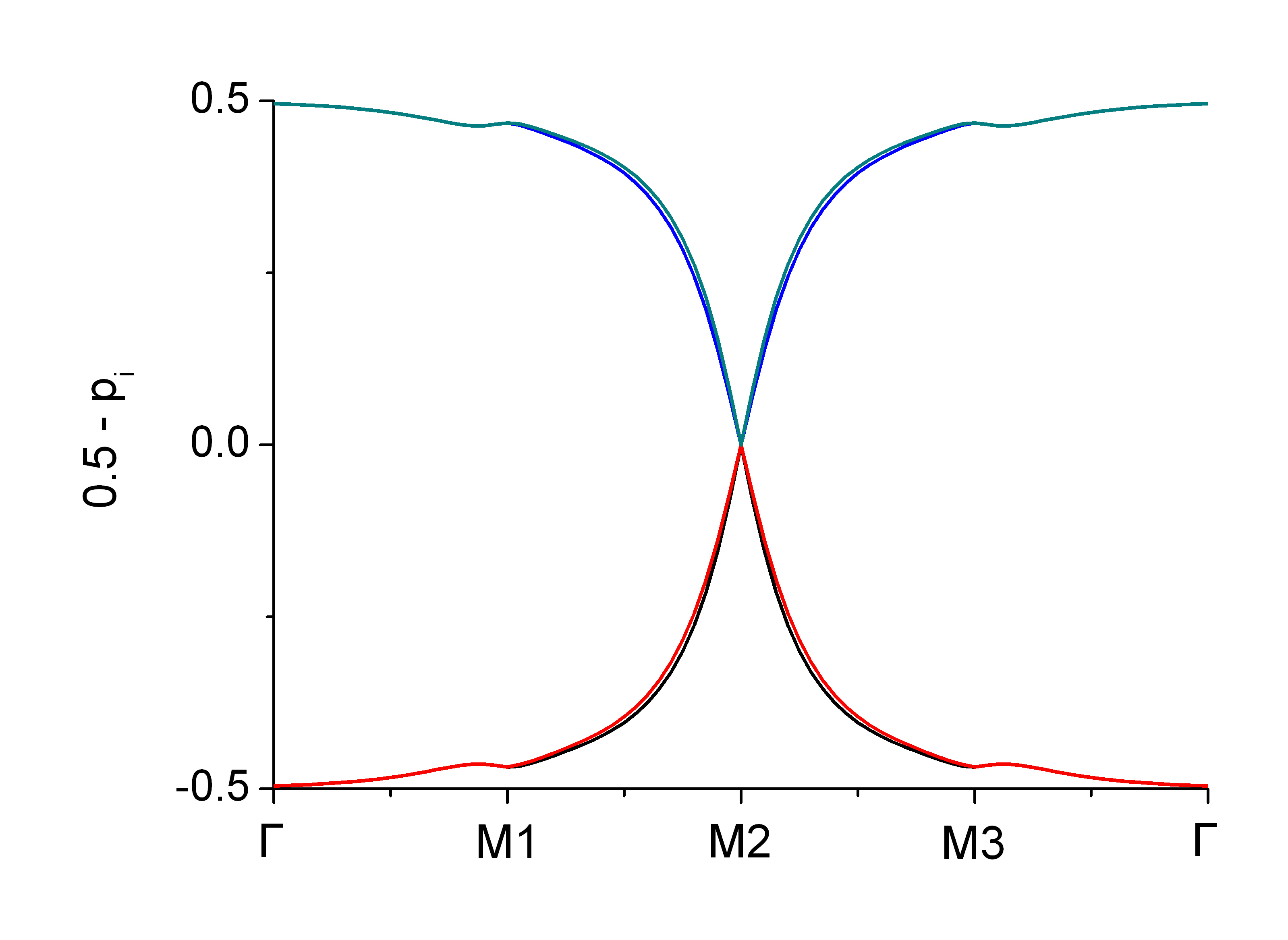}
} \caption{Inversion symmetric insulators with broken
time reversal symmetry. Note the physical surface
states (a) are gapped, but the Dirac node remains in the entanglement
spectrum (b).}
\label{fig:NoTwithI} 
\end{center}
\end{figure}

\begin{figure}[h!]
\begin{center}
\subfigure (a) 
{
    \label{fig:noIdispersion}
    \includegraphics[width=0.35\textwidth]{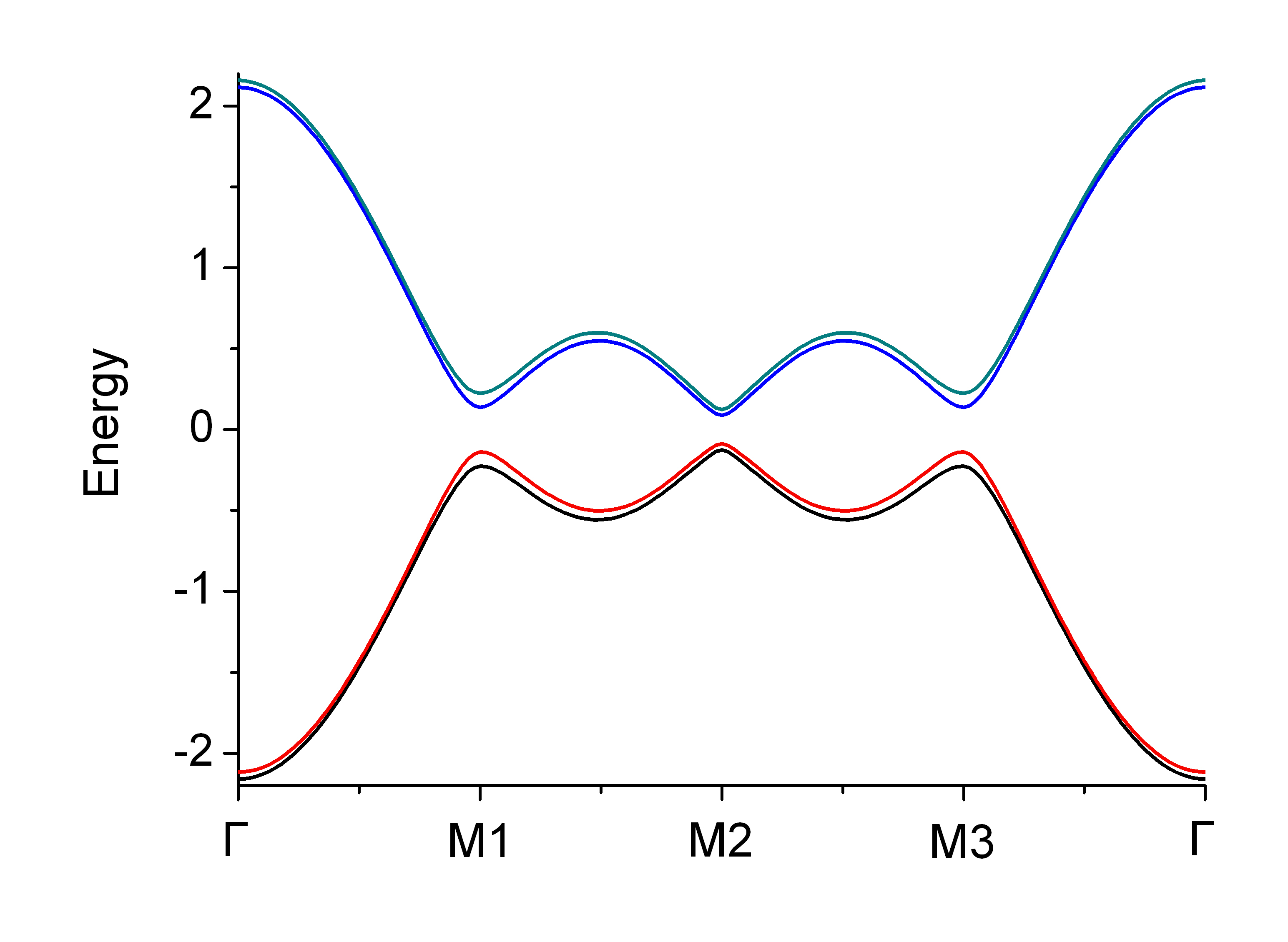}
} \hspace{2cm}
\subfigure (b) 
{
    \label{fig:noIspectrum}
    \includegraphics[width=0.35\textwidth]{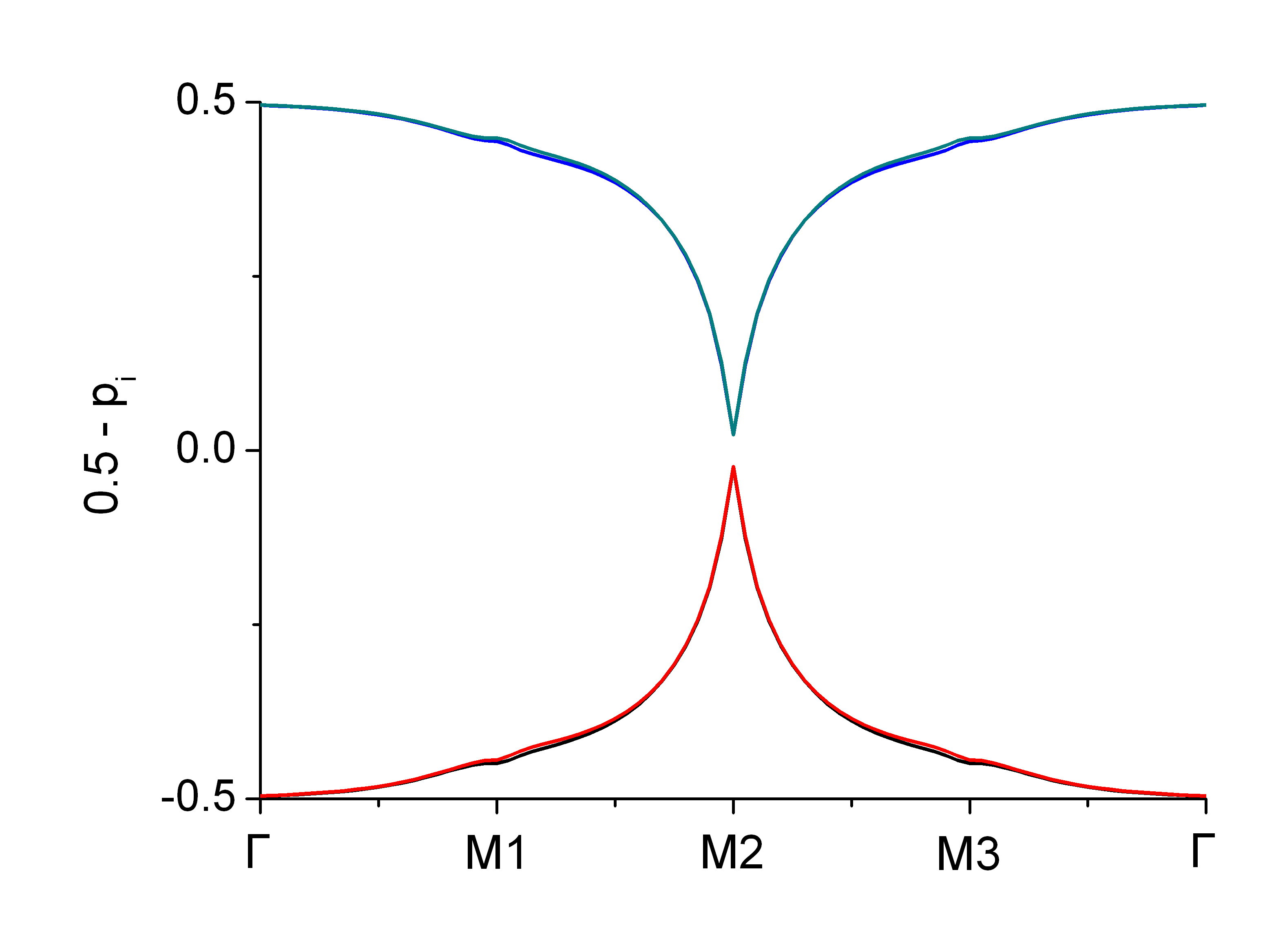}
} \caption{Without inversion symmetry. Both the physical surface
state (a) and the entanglement spectrum (b) are gapped. }
\label{fig:NoTnoI} 
\end{center}
\end{figure}

This has demonstrated that when a protected physical surface
state is present, this is reflected in the entanglement spectrum.
That the  Dirac nodes in the
entanglement spectrum are centered at zero, in contrast to the
physical surface spectrum in the same model, is an important observation.
Moreover, the entire entanglement spectrum remains symmetric under
$\epsilon^e\rightarrow -\epsilon^e$,
$\bf{k}_\perp\rightarrow-\bf{k}_\perp$. We will see that this is a
consequence of {\em inversion} symmetry, and the Dirac node in the
entanglement spectrum remains secure even when time reversal
invariance is broken. This is shown in Figure \ref{fig:NoTwithI}.
The physical spectrum of the surface states Fig. \ref{fig:NoTwithI}a
is gapped because of a uniform Zeeman field
$H_T=h\sum_ic_i^\dagger\sigma_z c_i$ applied to all the sites, with
$h=0.9$. However, the entanglement spectrum Dirac node remains
intact, despite the loss of time reversal symmetry in
Fig.\ref{fig:NoTwithI}b. When both time reversal symmetry and
inversion symmetry are destroyed, the latter via a staggered
potential $H_{I}= V\sum_i(-1)^ic^\dagger c_i$ (with $V=0.1$), then
the entanglement spectrum is also gapped as in Figure
\ref{fig:NoTnoI}b.

\section{Topological insulators with only Inversion Symmetry\label{spirit}}

We now discuss the origin of Fig. \ref{fig:NoTwithI}, where a
perturbation breaking time-reversal symmetry has been added. The
entanglement spectrum still has a Dirac node, which is in fact at
zero entanglement ``energy", although the surface states are gapped.
Therefore,
while a system with topologically protected
surface states also has topologically protected entanglement
states, the converse may not be true.
The Dirac node is
a feature in the
entanglement spectrum of inversion symmetric
insulators, which defines a distinct phase, although
there are no
topological surface states for the physical Hamiltonian
generically.
 The entanglement
spectrum is an especially useful tool for identifying insulators of this type.

The reason the entanglement spectrum works for identifying phases
with
inversion symmetry, while physical surface
states fail, is as follows.
A physical boundary necessarily breaks
inversion, since inversion maps the insulator to its
exterior.  But now consider a cut that passes through
a center of inversion. The inversion $\mathcal{I}$ through
this point
relates the left and right sides of
an entanglement cut. This inversion symmetry leads to
a symmetry of the
entanglement Hamiltonian, but with the twist that it
is realized as a sort of particle-hole symmetry for the two dimensional
entanglement spectrum (see also Refs.~ \onlinecite{frankA,frankB}).

Let us suppose that the inversion symmetric
cutting plane
does not exactly coincide with any orbitals(so that it divides the states into separate
parts)\footnote{We need to assume that, if the model
under consideration is a tight-binding model, then at least one center
of inversion symmetry does not coincide with a site.  Otherwise, there
is no inversion-symmetric way of dividing the sites into two parts.}.
We call the inversion transformation through the
center of inversion on the plane $\mathcal{I}$.
In this section,
we will describe how this symmetry constrains the entanglement states, 
then use this to
prove the stability of the entanglement Dirac node. Finally we
discuss what physical difference could be embodied in this
topologically distinct entanglement spectrum. The remaining sections
will discuss interacting insulators.

To see why inversion symmetry acts as a particle-hole symmetry
we will give an additional interpretation of the entanglement eigenstates
$f_{i\mathbf{k}_\perp}^R$.
The entanglement spectrum can be understood most intuitively by showing
that a set of
wave functions for the full system\cite{botero,klichsplit}:
\begin{multline}
F_{i\bf{k}_\perp}(x,\bf{r}_\perp)\\
=\left [ \sqrt{p_i(\bf{k}_\perp)}
f^R_{i\bf{k}_\perp}(x)+  \sqrt{1-p_i(\bf{k}_\perp)}
f^L_{i\bf{k}_\perp}(x)\right ] e^{i\bf{k}_\perp\cdot \bf{r}_\perp}
\label{Fff}
\end{multline}
can be constructed, satisfying two properties: first, the $f^R$'s
and $f^L$'s form an orthonormal family of wave-functions in the two
halves. Second, filling all the states $F_{i\bf{k}_\perp}$ gives the
band insulator; i.e. they can be thought of as being obtained from
the \emph{filled} band wavefunctions (Eq. \ref{bandwavefns}) by a
unitary transformation. The $f^R$'s and the $p_i$'s turn out to be
the same as before.  (See the appendix.) The eigenvalues
$p_i(\bf{k}_\perp)$ have a simple interpretation in light of Eq.
(\ref{Fff}). Each state $F_{i\bf{k}_\perp}$ is certainly occupied by
an electron in the band insulator. The eigenvalue
$p_i(\bf{k}_\perp)$ represents the probability that the electron in
this state is found on the right half. For example, when $p_i\approx 1$,
corresponding to $\epsilon_i\rightarrow-\infty$, the electron in
$F_i$ is almost certainly on the right half. This fits with the
previous definition of the $f^R$'s because $\epsilon_i$ is far below
the ``Fermi energy," so the $i^\mathrm{th}$ state is occupied in the
``Fermi sea" of the entanglement Hamiltonian. 
Eqn. (\ref{Fff}) leads to a very intuitive way of understanding
the entanglement spectrum,
as
reviewed in appendix A. 

\begin{figure}
\includegraphics[width=.45\textwidth]{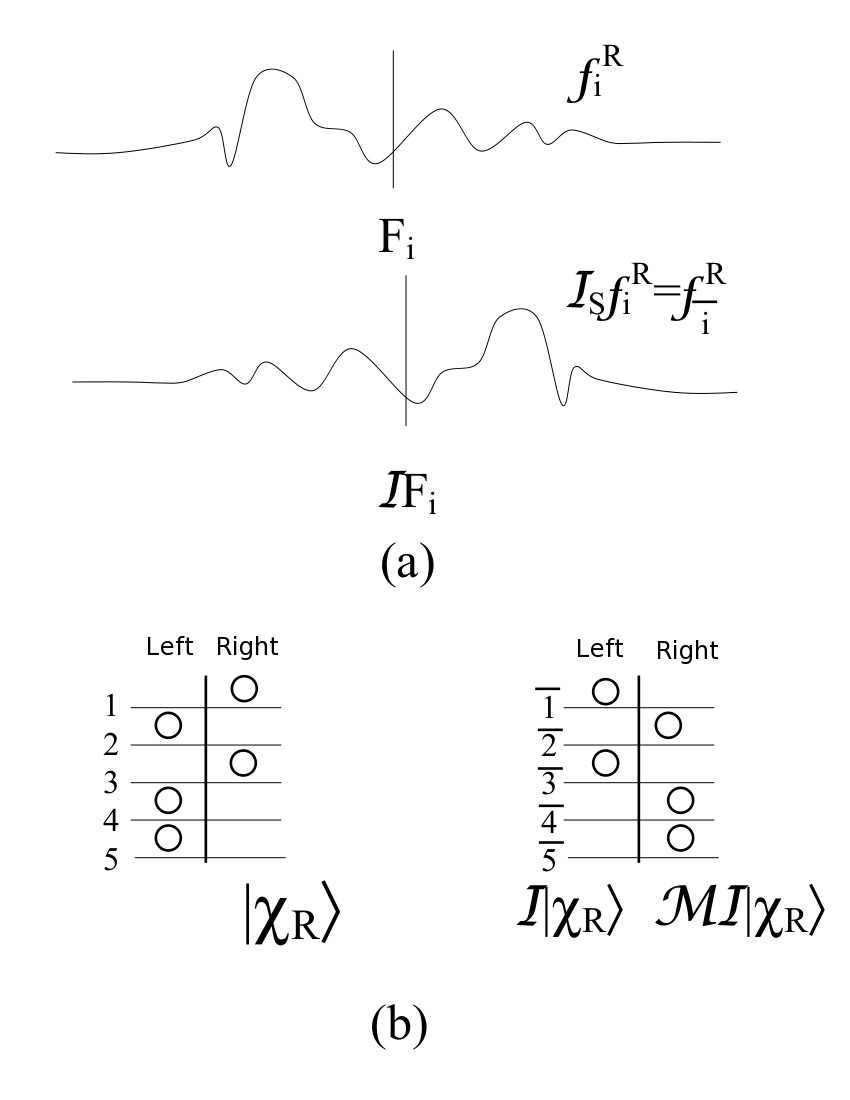}
\caption{Inversion symmetry of entanglement.
(a) The transformation $\mathcal{I}_S$ of single-body modes. The image of a state $f^R$ under
$\mathcal{I}_S$ is not defined on the merits of $f^R$ itself (like
ordinary reflection), but instead
depends on the state of the system.
  The \emph{extended}
wave function ($F_i=\sqrt{p_i}f^R_i + \sqrt{1-p_i}f^L_i$) is 
inverted, and $\mathcal{I}_Sf^R$ is (up to normalization) the
right-hand part of this, which can look completely different from
the original $f^R$. (b) The transformation $\mathcal{MI}$ of
many-body states in a non-interacting system. Each term in the
Schmidt decomposition is obtained by placing electrons on either the
right or left of the $F_i$'s indicated by the horizontal lines.  Inversion
is applied to the system as a whole.
Focusing just on the half right-hand side of the system, one finds that
inversion
induces a particle-hole like transformation ($\mathcal{MI}$) because levels
$i$ that are occupied in $|\chi_R\rangle$ correspond to
 empty levels $\bar{i}$ in $\mathcal{MI}|\chi_R\rangle$.
\label{soundtrack}
}
\end{figure}

Because of the inversion symmetry,
there is a map $\mathcal{I}_S$ on the \textbf{s}ingle-particle states that
takes a state $f_{i\mathbf{k}_\perp}^R$ to another state
$f_{\bar{i}-\mathbf{k}_\perp}^R$. This state flips the sign of the entanglement
energy, $\epsilon_{\bar{i}}^e(-\mathbf{k}_\perp)=
-\epsilon_i^e(\mathbf{k}_\perp)$. The mapping arises directly
from inversion symmetry acting on the $F$'s (see Fig. \ref{soundtrack}a),
which produces
from any occupied state $F_i$ with momentum $\mathbf{k}_\perp$ a second
state, $F_{\bar{i}}$, with
the opposite momentum:
\begin{equation}
F_{\bar{i}-\mathbf{k}_\perp}(x,\mathbf{r}_\perp)=F_{i\mathbf{k}_\perp}(-x,-\mathbf{r}_\perp).
\end{equation}
Expanding this equation gives
\begin{multline*}
\left [  \sqrt{p_{\bar{i}}(-\mathbf{k}_\perp)}
f^R_{\bar{i}-\mathbf{k}_\perp}(x)+  \sqrt{1-p_{\bar{i}}(-\mathbf{k}_\perp)}
f^L_{\bar{i}-\mathbf{k}_\perp}(x)\right]
e^{-i\mathbf{k}_\perp\cdot \mathbf{r}_\perp}
\\=\left [  \sqrt{p_i(\mathbf{k}_\perp)}
f^R_{i\mathbf{k}_\perp}(-x)+  \sqrt{1-p_i(\mathbf{k}_\perp)}
f^L_{i\mathbf{k}_\perp}(-x)\right]e^{-i\mathbf{k}_\perp\cdot \mathbf{r}_\perp}.
\end{multline*}
Since inversion maps the left-hand side to the right-hand side,
these equations imply that
$p_{\bar{i}}(-\mathbf{k}_\perp)=1-p_i(\mathbf{k}_\perp)$ and that
$f_{\bar{i}}^R$ is obtained from a state on the other side of the
partition, namely $f_{\bar{i}-\mathbf{k}_{\perp}}^R(x)
=f_{i\mathbf{k}_\perp}^L(-x)$. Using the relation between $p$ and
$\epsilon$, it follows that a mode with ``energy" $\epsilon^e$ and momentum
$\mathbf{k}_\perp$, is mapped by inversion to one with $-\epsilon^e$
and $-\mathbf{k}_\perp$. A more algebraic proof of this result is in
Appendix \ref{levi}.

Having established the action of inversion, we now turn to the
stability of Dirac modes in the entanglement spectrum, in the
presence of inversion. Suppose the insulator is obtained from a
 time-reversal symmetric topological insulator by applying a time reversal breaking perturbation. (We plan to
derive the basic properties of general inversion symmetric
insulators later\cite{future}.~\!\!) A Dirac mode in the entanglement
spectrum (or physical surface spectrum) occurs at a time reversal
invariant momentum (or TRIM) $\bm{\kappa}_\perp$.  These momenta are
half of a reciprocal lattice vector, i.e.,
$\bm{\kappa}_\perp\in\{\Gamma,\,M1,\,M2,\,M3\}$,
 so that
$\bm{\kappa}_\perp\equiv -\bm{\kappa}_\perp$
(modulo the reciprocal lattice).

First consider the modes in the unperturbed crystal, that is both
time-reversal and inversion symmetric.
At the tip of the cone of a Dirac mode, there are two degenerate
states $f^R_{a\bm{\kappa}_\perp},\ f^R_{b\bm{\kappa}_\perp}$. These must have $\epsilon^e=0$, because
otherwise applying $\mathcal{I}_S$ would produce a second Dirac point
at $-\epsilon^e$.

Now the symmetry $\mathcal{I}_S$ ensures that
these Dirac modes have to remain at zero ``energy"
when time-reversal symmetry is broken by a small perturbation.
First, the two states are transformed
into themselves under $\mathcal{I}_S$, so appropriate linear combinations of them
are inversion eigenstates.
In fact, their inversion parities must be the same.
(The states formed a Kramers doublet under time-reversal symmetry before
it was broken.) Therefore the states
$f^R_{a\bm{\kappa}_\perp}$ and $f^R_{b\bm{\kappa}_\perp}$
cannot evolve into a pair
$f^R_{+,\bm{\kappa}_\perp}$, $f^R_{-,\bm{\kappa}_\perp}$
with energies $\pm\epsilon^e$, without
a discontinuous jump in the inversion parities. (Inversion symmetry
interchanges these new states, so their combinations $\frac{1}{\sqrt{2}}
(f^R_{+,\bm{\kappa}_\perp}\pm f^R_{-,\bm{\kappa}_\perp})$ have
\emph{opposite} parities.)

Note, in general, the spectrum consists of equal numbers of states
that are even under inversion and odd under inversion. All the pairs
of states at \emph{nonzero} entanglement energy  can be combined
into pairs of states of opposite parity, since $f_{\epsilon,{\bf
k}_\perp}\rightarrow f_{-\epsilon,-{\bf k}_\perp} $, and one can
take symmetric and antisymmetric combinations of these. The only
exceptions are states at zero energy located at a TRIM. These are
actually the tips of Dirac nodes in the case at hand. Thus, $\Delta
\nu_{\bm{\kappa}_\perp}$, the difference between the number of even
and odd states at the TRIM $\bm{\kappa}_\perp$ is the number of
ungappable states.
Therefore, $\Delta \nu_{\bm{\kappa}_\perp}$ cannot change
except at a phase transition of the bulk crystal,
where the entanglement spectrum becomes ill-defined or changes discontinuously.
At a second order transition, in particular,
this topological invariance breaks
down because the entanglement Hamiltonian, Eqn. (\ref{Q}),
has long-range hopping and infinitely many bands (including
the bulk bands) collapse to zero energy (in analogy with
the observations of
Ref.~ \onlinecite{freeelec}). The states may then all mix together.
We will \emph{assume} here that the \emph{ungappable states all come in Dirac
pairs}, so $\Delta\nu_{\bm{\kappa}_\perp}$ is even.

The parity argument shows that there is a distinct phase of inversion symmetric
insulators defined by having a  protected Dirac dispersion in the entanglement
spectrum.
What physical property distinguishes this phase? Clearly, surface
states are not the answer since all physical surface states are gapped.
At least one distinction is captured by the electromagnetic
response of the system\cite{QiHughesZhang}. Consider integrating out
the Fermions in the presence of weak external electromagnetic
fields. Then the effective action contains a term:
\begin{equation}
S_\theta = i\theta \left [\frac{e^2}{2\pi h}\int d\tau d^3x {\bf
E}\cdot{\bf B} \right ].
\end{equation}
Inversion symmetry and time reversal symmetry each imply that $\theta$
is quantized in units of $\pi$.
Under time reversal,
the sign of this term is changed since ${\bf B}\rightarrow
-{\bf B}$. However, this does not rule out $S_\theta$. The reason
is that $S_{\theta}$ contributes
a factor of $e^{-S_\theta}$ to the weight of a field configuration in
the path integral.
The term in
square brackets is an integer for periodic boundary conditions
in space and imaginary time, so $\theta$ and
$\theta+2\pi$ are physically equivalent.  Hence effects
due to $S_{\theta}$ are time reversal invariant as long as $\theta$ is
0 or $\pi$. Note furthermore that if the system only possesses inversion
symmetry in the bulk, then one can argue instead that
the sign of this term changes
because ${\bf E}\rightarrow -{\bf E}$ under inversion.

The nonzero value, $\theta=\pi$, is realized in
strong topological insulators, time-reversal symmetric systems
with an odd number of surface Dirac nodes\cite{QiHughesZhang}.
It is therefore natural also to suppose that an insulator with an odd number
of entanglement nodes, but with only inversion symmetry,
has an electromagnetic response of
$\theta=\pi$. (This remains true even for inversion symmetric insulators
that cannot be obtained by perturbing time-reversal symmetric
insulators\cite{future}.~\!\!)
Note, the entanglement spectrum gives us a very simple way to
predict how the insulator \emph{responds} to
an electromagnetic field based solely on  \emph{ground state} properties.

The physical meaning of $S_\theta$ is that applying a magnetic
field induces a parallel polarization of charge
of magnitude $\theta \frac{e^2}{2\pi h}\mathbf{B}$.
Measuring this sharply is challenging, but it is at least in principle
a physical
consequence of the entanglement nodes.
The locations and numbers of nodes are also invariant, at least without
interactions, but we do not know the physical consequences of these properties.

\section{Stability against Interactions}
Thus far, we have discussed topological properties
of systems without interactions. An interesting question
is how many of the topological distinctions remain when the interactions
between electrons are taken into account. When surface states exist,
one can determine whether interactions affect their properties by studying
whether the interactions are ``relevant perturbations" to the field
theory of the Dirac modes\cite{key-2}.
Furthermore, the bulk
magnetoelectric polarizability remains quantized even when interactions are
included. But the entanglement spectrum remains gapless even when there
are no surface states and captures quantum numbers not accounted
for by $\theta$:  at the very least, we believe the values of
$\Delta\nu_{\bm{\kappa}_\perp}$ are conserved modulo 2 at each transverse
TRIM. (Similar invariants can be constructed for each direction
of the entanglement cut, but presumably only a few of these are
independent.)

To describe inversion symmetry in an interacting state, one must
understand how it acts on the {\em many body} states appearing in
the Schmidt decomposition.
Inversion turns out to be related to
an {\em anti-unitary} operator, $|\chi^R\rangle\rightarrow
\mathcal{MI}|\chi^{R}\rangle$.
The action of this symmetry is most interesting when
restricted to the even-split Schmidt states, where half the $N$
particles are on each side of the divider.
Here, it satisfies $(\mathcal{MI})^2=-1$
when there are an odd number of Dirac nodes.
Thus, inversion behaves exactly like the Kramers transformation, and
each Schmidt state is doubly degenerate even with interactions included.
Interestingly, Fermion {\em anticommutation} is a key ingredient in
establishing this fact. It has been shown recently\cite{frankA,
frankB} that inversion acts on one dimensional Haldane chains in a
similar way.

\subsection{Many Body States and Inversion Symmetry}
Consider the action of inversion symmetry $\mathcal{I}$ on the many body Schmidt
states \begin{equation}|\Psi_G\rangle = \sum_a
e^{-E^e_a/2}|\Phi^R_a\rangle|\Phi^L_a\rangle\label{Schmidt}.\end{equation}
Inversion maps a right-hand state to a left-hand one.
Since
this is a symmetry, the two must have the same entanglement
eigenvalue $E_a^e$. Thus, if each eigenvalue has a non-degenerate
eigenstate, the inversion transformation is simple: each state
maps to its partner (up to a phase). Things get
more interesting when degenerate states are present in the entanglement
spectrum.

Consider a multiplet of even-split states
$|\Phi^R_a\rangle$, where we use the label $a=1,\dots, d$ to label the $d$
degenerate states in the Schmidt basis above, Eq. (\ref{Schmidt}).
Its image under
inversion is a linear combination of states on the
left:
\begin{equation}
|\Phi^R_a\rangle \rightarrow \sum_{b=1}^N I_{ba} |\Phi^L_b\rangle
\end{equation}
If inversion is to be viewed as a symmetry, we need to return to the
right portion of the system, to make statements about the
entanglement eigenstates of a single subsystems. (See Fig. \ref{soundtrack}b.)
(Symmetries mapping systems to one another are not so useful--for example
knowing that the mirror image of a left-handed molecule is
a right-handed molecule does not imply that the orbitals associated
with
one of these molecules have special symmetry.)
Inversion can be converted to a symmetry of
one part using the pairing of left and right states in the
Schmidt decomposition (\ref{Schmidt}). This pairing is
described by a transformation
$M$ which maps $|\Phi_a^L\rangle\rightarrow |\Phi_a^R\rangle$.
This rule
can be extended consistently to an \emph{antilinear} transformation on
the rest of the Hilbert space for the left
side of the system.

%

To see this, let us describe the
partner of a general state $|\chi^L\rangle$ on the left
in a basis-independent way:
\begin{equation}
\mathcal{M}|\chi^L\rangle=\langle\chi^L | {\rho_R}^{-\frac12} |\Psi_G \rangle.
\label{match}
\end{equation}
The right-hand side is a partial inner product. It is not a number
but a \emph{wave function} for the right half of the system since
only the degrees of freedom on the left half are summed over:
$\mathcal{M}|\chi_L\rangle=\sum_{\{a^L_i\}}
\Psi_G(\{a^R_i\},\,\{a^L_i\})\chi^{L*}(\{a^L_i\})\,|\{a_i^R\}\rangle$
where $\{a^L_i\},\{a^R_i\}$ are the variables describing the two
halves.
Note that this is antilinear in $|\chi^L\rangle$
because of the complex conjugation. Now,
$|\chi^L\rangle=\sum_ac_a|\Phi_a^L\rangle$
is mapped by this transformation to $\sum_i c_a^*|\Phi_a^R\rangle$.
Each Schmidt state, in particular, maps to its partner.
Briefly, Eq. (\ref{match}) utilizes the ground state to connect the left
and right halves, and the
operator ${\rho_R}^{-\frac12}$ is there
to strip off the different
Schmidt weights.

Now the combination of $\mathcal{M}$ and $\mathcal{I}$ is a stand-in for
inversion
symmetry that returns states on the right-hand side of the
cut back to the right-hand side; in the basis of
Schmidt states, this transformation is represented by
$KI$ where $K$ is complex conjugation and $I$ is the matrix given above.

When $\mathcal{MI}$ is performed twice, the state
must return to itself, modulo a phase, hence
\begin{equation}I I^* = {\mathbf 1}e^{i\phi}.\end{equation}
This
phase factor can be only $\pm 1$ because it has to be real\footnote{
 The relationship  implies
$I^*=e^{i\phi}I^{-1}$ (note, we can take the inverse since
the determinant of $I$ is nonzero on account of the initial
relationship).  Now multiplying by $I$ on the left gives back
$I\cdot I^* ={\mathbf 1}e^{i\phi}$.  On the other hand,
taking the complex conjugate of the original expression
gives $I\cdot I^* ={\mathbf 1}e^{-i\phi}$ so
$\phi=0, \,\pi$}.
More precisely,
the operator $(\mathcal{MI})^2$ has to have the same value $(-1)^\delta$ for
each Schmidt state which has half the electrons on each side,
since the wave function Eq. (\ref{schmidt}) is a parity
eigenstate (see Appendix \ref{pion}).

The case $I\cdot I^* = -{\mathbf 1}$ is specially interesting.
Time reversal symmetry of spin $1/2+n$ particles has the same property, which
can be used to prove Kramers' degeneracy.
This algebra has no one dimensional representation. (If
$I$ is a c-number, the product is nonnegative.)
A two dimensional
representation is exemplified by
                                  $I=\left (\begin{array}{cc}
                                    0 & -1 \\
                                    1 & 0
                                  \end{array}\right )$, the antisymmetric
                                  matrix. Since this is actually the
                                  generic case, whenever the
                                  inversion operator in a particular
                                  system obeys this
                                  algebra, all levels must be at
                                  least doubly degenerate.

We now show that this is indeed how
the inversion operator acts in the even-split states 
when there is an odd number of Dirac nodes,
by showing that $\delta$ is the number of Dirac nodes modulo 2.
For states $|\Phi_R\rangle$ with $\frac{N}{2}+k$ Fermions
on the right side, $(\mathcal{MI})^2$ can be different, but
there will still be entanglement degeneracies if $k\neq 0$:
the entanglement eigenvalues for positive
and negative $k$ always match, since
the Schmidt state $|\Phi_R\rangle$ is
degenerate with $|I\Phi_L\rangle$ (and has the opposite $k$).
This relation (valid for any inversion-symmetric
insulator) is not as interesting as the degeneracies
among the states with the same value $k=0$, which is unique
to topological insulators.



\subsection{Inversion and the Topological Insulator}
We now determine the value $II^*=\pm 1$ for topological insulators
in the noninteracting limit. Note, this also allows us to make
statements about the interacting case, since the value cannot jump
when small interactions are introduced.

The ground state of the noninteracting system
can be built up by creating particles in
all the modes $F_i$ (see Eq. (\ref{Fff})). Suppose there is
just one Dirac node. Then
\begin{alignat}{1}
&|\Psi_G\rangle = \frac12 (r_a^{\dagger}+l_a^{\dagger})(r_b^{\dagger}
+l_b^{\dagger})\nonumber\\
&\ \
\times\prod_{\begin{subarray}{c}i\mathbf{k}_\perp\\ p_i(\mathbf{k}_\perp)>\frac12
\end{subarray}}
(\sqrt{p_i(\mathbf{k}_\perp)}r^\dagger_{i\mathbf{k}_\perp} +
\sqrt{1-p_i(\mathbf{k}_\perp)}l^\dagger_{i\mathbf{k}_\perp})\nonumber\\
&\ \
\times\prod_{\begin{subarray}{c}i\mathbf{k}_\perp\\ p_i(\mathbf{k}_\perp)<\frac12
\end{subarray}}
(\sqrt{p_i(\mathbf{k}_\perp)}r^\dagger_{i\mathbf{k}_\perp} +
\sqrt{1-p_i(\mathbf{k}_\perp)}l^\dagger_{i\mathbf{k}_\perp})
|0\rangle \label{expand}\\
&r_{i\mathbf{k}_\perp}^\dagger
=\sum_{{\bf r}} e^{i{\bf k}_\perp\cdot \mathbf{r}_\perp}
f^R_{i{\bf k}_\perp}({\bf r})\psi(\mathbf{r})^\dagger\\
&l_{i\mathbf{k}_\perp}^\dagger
=\sum_{{\bf r}} e^{i{\bf k}_\perp\cdot \mathbf{r}_\perp}
f^L_{i{\bf k}_\perp}({\bf r})\psi(\mathbf{r})^\dagger.
\end{alignat}
where the first two factors are the Dirac node states. Expanding Eq.
(\ref{expand}) gives the Schmidt decomposition. One should convince
oneself of the following relation between the many-body transformations
$\mathcal{MI}$ and the one body implementation of inversion,
$\mathcal{I}_S$: if $|\chi_R\rangle$ is one of the Schmidt states,
and $f^R_i$ is occupied in this state, then $\mathcal{I}_Sf^R_i$ is
empty in $\mathcal{MI}|\chi_R\rangle$ (see Fig. \ref{soundtrack}b).

Note that
the second line of Eq. (\ref{expand})
contains the states that are mostly on the right,
and the third contains those mostly on the left. Inversion maps these
states to one another. The two states at the Dirac
node are inversion eigenstates,
 so they map to themselves.
Since they also have the same parity (say they are even,
for instance),
$r_{a,b}\leftrightarrow l_{a,b}$.

The highest weight states in the Schmidt decomposition
 of the wavefunction involve acting with
$r^\dagger_i$'s when $p_i>1/2$ and with $l^\dagger_i$'s
when $p_i<1/2$, so these states are contained in
\begin{equation}
\frac12(r^\dagger_a +
l^\dagger_a)(r^\dagger_b+l^\dagger_b)|S_R\rangle |S_L \rangle
\label{twoseas}
\end{equation}
where $|S_R\rangle$ and $|S_L\rangle$ are the filled Fermi seas (all
negative single particle entanglement energies occupied) for the two
sides, which are exchanged by $\mathcal{I}$.

Among the highest weight states, consider the two states
$$
|pair\rangle=r^\dagger_al^\dagger_b|S_R\rangle |S_L \rangle +
l^\dagger_ar^\dagger_b|S_R\rangle |S_L \rangle
$$
with an equal number of Fermions in the two sides.
While these are converted into each other under inversion, it
seems possible that they could mix and split, and give rise to
one dimensional representations of the inversion operation. This is
where showing that the inversion matrix $I$ satisfies $I
I^*=-{\mathbf 1}$ comes in handy.

We can get $|pair\rangle$ into the form of the Schmidt decomposition
if we
define the states:
\begin{eqnarray*}
|\Phi^R_1\rangle = r^\dagger_a |S_R\rangle && |\Phi^L_1\rangle =
l^\dagger_b |S_L\rangle\\
|\Phi^R_2\rangle = -r^\dagger_b |S_R\rangle && |\Phi^L_2\rangle =
l^\dagger_a |S_L\rangle.
\end{eqnarray*}
The ground state can be written as the Schmidt sum:
$$
|\Psi_0 \rangle =s\frac12 (|\Phi^R_1\rangle|\Phi^L_1\rangle+
|\Phi^R_2\rangle|\Phi^L_2\rangle)+\dots
$$
(The sign $s$ in front is an unimportant sign due to Fermi statistics;
see appendix \ref{pion}.)

Inversion maps
\begin{eqnarray*}
|\Phi^R_1\rangle \mapsto |\Phi^L_2\rangle \\
|\Phi^R_2\rangle \mapsto -|\Phi^L_1\rangle.
\end{eqnarray*}
 $\mathcal{M}$ simply maps $|\Phi^L_{1,2}\rangle$ to
$s|\Phi^R_{1,2}\rangle$, so the inversion matrix is
$I^*=\left(\begin{array}{cc}0&-s\\s& 0\end{array}\right)$.
Note that the crucial minus sign has arisen 
because of anticommutation of Fermion
operators.

Now we see explicitly that the inversion matrix satisfies $I\cdot
I^*=-{\mathbf 1}$, and hence the states remain two fold degenerate.
As mentioned above, this result persists for any state with
equal numbers of Fermions on the two sides. When interactions are included, 
the even-split states all
mix together, but $I$ can be enlarged to describe the action of inversion on
the whole space. 
$I\cdot I^*$ remains equal to
$-{\mathbf 1}$, so the degeneracies survive.
Unlike the usual Kramers
degeneracy for time-reversal, this result does not require an odd
number of Fermions.

For a more general topological insulator, one can show
that $I\cdot I^*=(-)^\delta{\mathbf 1}$ where
$\delta$ is the number of pairs of equal-parity single-body states
with $\epsilon^e=0$, i.e. the number of Dirac nodes.





\subsection{Distinctions Preserved by Interactions}
Now we can argue that some distinctions among insulators survive the
introduction of interactions. While this will include the quantized
electromagnetic response of inversion symmetric insulators
\cite{QiHughesZhang}, which by virtue of being a response function
remains well defined in the interacting case, other more mysterious
distinctions are also found.
Without interactions, $\frac{1}{2}\Delta \nu_{\bm{\kappa}_\perp}$ is a fixed
integer for each TRIM. With interactions, we will  give an argument
that suggests that at least the number of Dirac nodes at each
$\bm{\kappa}$, $\frac{1}{2}\Delta\nu_{\mathbf{\kappa}_\perp}$,
is well-defined modulo 2 in an inversion symmetric
insulator. In particular, the location of a single surface Dirac
node in the Brillouin zone survives the introduction of
interactions, as argued below. The analogous quantity in the time
reversal invariant insulator is the ``weak" index,
\cite{key-2,MooreBalents}, which can be determined from the surface
states \cite{key-2} or metallic topological defects in the crystal
\cite{Ying}. No analogous physical consequence seems available when
time reversal is broken but inversion is retained.

These distinctions can be found by looking at a sample with
a finite cross-section in the $y-z$ direction (but infinite
in the $x$-direction, perpendicular to the cut).
We have just seen that, in a noninteracting insulator,
the parities of the zero-``energy" single-body states
determine whether there are many-body degeneracies.
Only the Dirac modes at certain
TRIMs will satisfy
the boundary
conditions, and thus the value of $(\mathcal{MI})^2$ will count the number
of nodes at these TRIMs.

Say the cross-section is odd$\times$odd, with
$2N_y+1\times2N_z+1$ unit cells.  Assuming periodic boundary
conditions, the allowed
transverse momenta will be
$(\frac{2\pi n_y}{(2N_y+1)},\frac{2\pi n_z}{(2N_z+1)})$
(where $n_y,\ n_z$ are integers whose magnitudes are
less than or equal to $N_y,\ N_z$
respectively). The only one of these which is exactly time-reversal
invariant is
$(0,0)$. Thus all states come in pairs related by inversion
symmetry except for unpaired states $\frac{1}{\sqrt{2}}(l_{a,b}^\dagger
+r_{a,b}^\dagger)$ at zero momentum.
Therefore $\delta$ is equal to $\frac12\Delta\nu_{(0,0)}$ and there
is a double degeneracy if this is odd.

 To isolate another TRIM,
introduce antiperiodic boundary conditions along one or
both of the other directions. (Antiperiodic boundary conditions
along $y$, for example, force $k_{y\perp}$ to have the form
$\frac{2\pi (n_y+\frac{1}{2})}{(2N_y+1)}$, allowing $\pi$ but not
$0$.) Then double degeneracy occurs when the number of
modes at the new TRIM is odd.

It is possible that the actual integer value of the $\Delta \nu$'s
is conserved also when interactions are introduced, but there might
be a more surprising classification of interacting phases. Refs.
\cite{Kitaev,LukaszKitaev} gave an example for
\emph{one}-dimensional interacting topological insulators showing
that an integer property of topological phases can be changed (by
multiples of 8, in fact) when interactions are included.

\subsection{The Parity of the Many-body Wavefunction}
Many phases (like the ordered phase of an antiferromagnet) have
a sharp distinction only for infinitely large systems, but
it is possible to check what phase an inversion-symmetric insulator is
by looking at a finite piece of it, with an appropriate geometry.

For a sample with periodic boundary conditions in the $y$ and $z$
directions, a finite size in the $x$-direction, and perfect inversion symmetry
between its two free surfaces,
$\frac{1}{2}\Delta \nu_{(0,0)}+\frac{N}{2}$ is
the parity of the many-body wave-function $\psi(\bm{r}_1,\dots,\bm{r}_N)$
under $\mathbf{r}\rightarrow -\mathbf{r}$.
This
can be seen when $\Delta\nu_{(0,0)}=1$
by inverting Eq. (\ref{twoseas}): the two modes corresponding
to the Dirac node just
map to themselves under inversion symmetry while $|S_R\rangle$ and
$|S_L\rangle$ switch places. Switching them back leads to a factor of
$(-1)^{N/2-1}$ because $N/2-1$ pairs of electrons have been
exchanged. (See Appendix \ref{pion} for more details.)
This parity is not any easier to measure experimentally than degeneracies in
the entanglement spectrum; a measurement would require some sort
of interference experiment on a macroscopic crystal\footnote{One could
send the crystal toward a beam-splitting ``mirror" which inverts the
crystal with probability one-half, and then interfere the beams.  Making a mirror that can turn the crystal
inside-out like this is even harder than ``just" maintaining
the coherence of a macroscopic crystal during an interference
experiment!}; it is just an interesting
interpretation for $\Delta \nu_{(0,0)}$. (The $\Delta\nu$'s at other
TRIMs are related to the parities of samples with other boundary
conditions. )

To check what phase
a sample is in if it is \emph{not} inversion symmetric, the system
does have to be very large. In this case, one has to use the
entanglement spectrum to determine the phase, and this
works only when the surfaces are far enough from the cut surface that
inversion is an accurate symmetry of the entanglement surface states.

As an aside, this implies a surprising relationship between the
magnetoelectric susceptibility and the parity when inversion is a
symmetry: the electron-wave function of a crystal with a
half-integer susceptibility is odd under inversion for the following
geometry: the sample must have an even$\times$even cross section and
periodic boundary conditions in the $y-z$ direction. (For the
even$\times$even cross sections all the TRIMs contribute to the
parity).



\section{Conclusions}
We have studied the entanglement spectrum of insulators with non-trivial
band topology. Whenever a physical edge or surface
state is present, the entanglement spectrum also is characterized by
protected features. Although this is purely a ground state property,
we showed it can be formally mapped to the edge spectrum of a ``flat
band" version of the physical Hamiltonian.

On the other hand, the converse of this is not true. Protected
features of the entanglement spectrum may occur in systems that do
not possess physical surface states. This can occur when a physical
surface necessarily breaks an underlying symmetry of the bulk solid
- for example inversion symmetry. In these cases the entanglement
spectrum provides a unique perspective on classifying the phase. We
illustrated this by studying three dimensional insulators with
inversion symmetry. In general, no surface modes occur in these
systems. However, since the entanglement cut still preserves
inversion symmetry, the action of inversion on the entanglement
eigenmodes can be worked out. These were shown to lead to the
protected entanglement spectrum.

An advantage of defining the phase in terms of protected properties
of its entanglement spectrum is that it allows us to deduce
properties that remain stable when interactions are present. For the
case of inversion symmetric insulators, characteristic properties beyond
the quantized magnetoelectric polarizability appear to be present and
are stable when the particles are interacting.  The corresponding
physical consequences remain to be identified.


\section{Acknowledgements}
AMT thanks Frank Pollman, Erez Berg, and Masaki
Oshikawa for a collaboration on spin chains that
helped us to understand the interacting case.
We also acknowledge support from NSF-DMR-0645691 is acknowledged.
\appendix
\section{Inversion Symmetry in Noninteracting Entanglement Spectra\label{levi}}
The transformation $\mathcal{I}_S$
$f^R_{i, \bm{\kappa}_\perp}\rightarrow f^R_{\bar{i},-\bm{\kappa}_\perp}$
can be expressed in terms of the correlation function $\hat{C}$.
First, we will show how to construct the paired functions 
$F_{i\bf{k}_\perp}(x)$.
(We will omit the $y$ and $z$ dependence.)
Let us split the wave function into two parts,
$x>0$ and $x<0$, which can be regarded as the top and bottom halves
of state-vectors.  Then the correlation function has four parts,
\begin{equation}
\hat{C}=\left(\begin{array}{cc}\hat{C}_R & \hat{C}_{LR}\\ \hat{C}_{LR}^\dagger &\hat{C}_L \end{array}\right).
\end{equation}
Since $\hat{C}$ has $1$ and $0$ as eigenvalues,
$\hat{C}^2=\hat{C}$, giving four matrix equations. The equation of
relevance is $\hat{C}_{LR}({\bf 1}-\hat{C}_R)=\hat{C}_L\hat{C}_{LR}$.
Given an eigenfunction $f^R_i$ of $\hat{C}_R$ with eigenvalue $p_i$
one can obtain an eigenvector of $\hat{C}_R$ with eigenvalue
$1-p_i$ via the transformation $\hat{M}$
\begin{equation}
f^L_i(x)=[\hat{M}f^R_i](x)=\frac{1}{\sqrt{p_i(1-p_i)}}
\sum_{x'>0} \hat{C}_{LR}(x,x')\phi_i^R(x').
\end{equation}
The prefactor is inserted to ensure that $f^L_i$ is normalized.
(One can check that $\hat{M}$ is a unitary transformation,
which can be written in matrix form
$\hat{M}=\frac{1}{\sqrt{\hat{C}_L-\hat{C}_L^2}}\hat{C}_{LR}$).

Next one can combine $f^L_i$ and $f^R_i$ to give wave functions
for the unpartitioned system.  One can show using the
relations between the submatrices of $\hat{C}$ that Eqn. (\ref{Fff}) is
an eigenstate of $\hat{C}$ with eigenvalue $1$, and hence is occupied.
Now the eigenstates of $\hat{C}_{L}$ and $\hat{C}_R$ are the entanglement
modes for the left and right sides of the systems.
Hence $F$ pairs up all the entanglement modes
into occupied states. (There are two exceptions: if $p_i=0$ or $1$,
then $M$ is not well-defined.  In the first case, $f^R_i$ is definitely
unoccupied, so there is
no $F_i$ corresponding to it.
In the second case, $f^R_i$ is definitely occupied and does
not require a partner.)

The expression for the Schmidt weights has an
intuitive relationship to the $F$'s. Since
each $F$ has its own electron with probability one,
a term
in the Schmidt decomposition is obtained (see Fig. \ref{soundtrack}b)
when a decision is made
about which of these electrons are to reside on the right, say those
in states $F_j$ with $j\in A_R$, and which are to reside on the
left (those in states with $j\in A_L$). 
The Schmidt coefficients
$\lambda_\alpha^2$ are thus given by a Bernoulli distribution,
$\prod_{j\in A_R}p_j\prod_{j\in A_L}(1-p_j)$, which is equivalent to
Eq. (\ref{eigenvalues}). 
(Formally, one obtains this result 
by expanding 
$|\Psi_G\rangle=
\prod_{i\mathbf{k}_\perp}\left(\sqrt{p_i(\mathbf{k}_\perp)}
r_{i\mathbf{k}_\perp}^\dagger+
\sqrt{1-p_i(\mathbf{k}_\perp)}l_{i\mathbf{k}_\perp}^\dagger\right)|0\rangle$.) 

Since $\hat{M}$ maps the right-half to the left half of the
system, it can be combined with inversion symmetry, $\hat{I}$, to
give the transformation used in Section VI,
which transforms the entanglement spectrum of the
right half into itself,  $\mathcal{I}_S=\hat{I}\hat{M}$. Since it
anticommutes with $\hat{C}_R-\frac{1}{2}$, $\hat{I}\hat{M}$
changes the sign of $p-\frac{1}{2}$ and hence of $\epsilon^e$,
as well as the sign of $\bf{k}_\perp$.  It is also unitary.
Thus, this transformation has the same properties as a
two-dimensional ``CRT" symmetry of the space parallel
to the cut. Here $C$ is a particle-hole
symmetry, $R$ is a $180^\circ$ rotation and $T$ is time-reversal
symmetry, although the system does not have those
symmetries independently. The $C$ factor changes the sign of the energy,
and all three factors change the sign of the momentum.
The product of all three is unitary, like $\mathcal{I}_S$,
because particle-hole symmetry is antiunitary
when one considers how it acts on \emph{single-particle}
states. 

This fact is enlightening since a $CPT$
symmetry (with a reflection $P$ in place of rotation)
could not be used to prove the
masslessness of the Dirac excitations: any relativistic equation
(including the massive Dirac equation)
is invariant under CPT symmetry.  
(In two dimensions, inversion through the origin is not $P$ symmetry,
since it is orientation-preserving, and equal to a $180^\circ$ rotation.)

\section{Parity of Finite Systems\label{pion}}

The result that $(\mathcal{IM})^2$ is a function only of
the topological phase of an insulator (characterized by $\delta$)
and of $k$, the number of excess particles in a Schmidt
state, can be proved most easily by considering
the inversion transformation of the ground
state wave function of a finite system. This wave function
must be an inversion eigenstate with
some parity $(-1)^P$.

Notice by the way an important point that we have not discussed much:
The products appearing in the Schmidt decomposition,
Eq. (\ref{schmidt}), have to be antisymmetrized:
\begin{alignat}{1}
|\alpha R\rangle|\alpha L\rangle\equiv A.S.\{\phi_{\alpha R}
&(\mathbf{r}_1,\mathbf{r}_2,\dots,\mathbf{r}_{\frac{N}{2}+k})\\
&\phi_{\alpha L}
(\mathbf{r}_{\frac{N}{2}+k+1},\dots,\mathbf{r}_N)\},\label{eq:overshot}
\end{alignat}
where ``$A.S.$" means to antisymmetrize in all the variables.
This wave function is not a product wave function, but it is the
closest thing possible for Fermions: The correlations between
densities on opposite sides of the cut vanish, surprisingly maybe\footnote{
Whether the two sides of the system are correlated or not on this
wave function depends on whether one is looking at an ``Eulerian
representation," (as it is called in the context of fluid mechanics) 
with variables
that are localized in space (such as particle number) or
a ``Lagrangian representation," with variables
that are attached to specific particles (their positions, for example). 
In the Lagrangian
representation, the variables $\mathbf{r}_1$ and $\mathbf{r}_2$
are entangled.}.


Let us call the antisymmetrized-product wave function $\Phi_\alpha$.
The inverse of this wave function is
\begin{alignat}{1}
\Phi_\alpha&(-\mathbf{r}_i)\nonumber\\
&=A.S.\ \{\Phi_{\alpha R}(-\mathbf{r}_1,\dots,
-\mathbf{r}_{\frac{N}{2}+k})
\Phi_{\alpha L}(-\mathbf{r}_{\frac{N}{2}+k+1},\dots,-\mathbf{r}_N)\}\nonumber
\end{alignat}
The inversion of the factor corresponding to the right side of the system
$\mathcal{I}\Phi_{\alpha R}=\Phi_{\alpha R}(-\mathbf{r}_i)$
is a wave function on the left side. So we should compare this
expression to
$|\mathcal{I}\Phi_{\alpha L}\rangle|\mathcal{I}\Phi_{\alpha R}\rangle$,
which is equal to
$A.S.\ \{\Phi_{\alpha_L}(-\mathbf{r}_1,\dots,-\mathbf{r}_{\frac N2-k})\Phi_{\alpha R}(-\mathbf{r}_{\frac{N}{2}-k+1},\dots,-\mathbf{r}_N)\}$. This differs
from $\Phi_\alpha(-\mathbf{r}_i)$ in the labelling of the coordinates.
The parity of the permutation is $(-1)^{(\frac{N}{2}-k)(\frac{N}{2}+k)}$
which is equal to
$(-1)^{\frac{N}{2}-k}$ and so
\begin{equation}
\mathcal{I}|\Phi_{\alpha R}\rangle|\Phi_{\alpha L}\rangle=(-1)^{\frac N2-k}
|\mathcal{I}\Phi_{\alpha L}\rangle|\mathcal{I}\Phi_{\alpha R}\rangle
\label{determinanty}
\end{equation}

Now let us calculate the parity of the ground state of a noninteracting
 insulator by looking at a particular term in the Schmidt decomposition.
Assume as we have been doing that there are
$\delta$ pairs of zero energy states
$\frac{1}{\sqrt{2}}(r_{ia}^\dagger+l_{ia}^\dagger)$
and $\frac{1}{\sqrt{2}}(r_{ib}^\dagger+l_{ib}^\dagger)$.  Since
each pair has the same parity $\pi_i=\pm 1$,
$r_{ia,b}\leftrightarrow \pi_i l_{ia,b}$.
Then all the maximum weight states of the
Schmidt decomposition
are contained in $|M.W.\rangle=\prod_{i=1}^{\delta} (r_{2i-1}^\dagger+\pi_il_{2i-1}^\dagger)(r_{2i}^\dagger+\pi_il_{2i}^\dagger)|S_R\rangle|S_L\rangle$.
By Eq. (\ref{determinanty}),
the parity of $|S_R\rangle|S_L\rangle$ is $(-1)^{\frac{N-2\delta}{2}}$.
The product of all the Dirac mode operators is even
under inversion, so the parity of $|M.W.\rangle$ is also
$(-1)^{\frac{N-2\delta}{2}}$.  This also has to be the parity $(-1)^P$
of the entire ground state wave function, as was claimed in Section
\ref{pion}.

Now consider a generic term of the Schmidt decomposition of $|\Psi_G\rangle$,
such as $|\Phi_\alpha\rangle$.  Applying inversion to the Schmidt
decomposition maps each term to another term
except for a factor of $(-1)^P$.
This is the
case for $|\Phi_\alpha\rangle=|\Phi_{\alpha R}\rangle|\psi_{\alpha L}\rangle$, 
so
$(-1)^{P+\frac{N}{2}-k}|\mathcal{I}\Phi_{\alpha L}\rangle|\mathcal{I}\Phi_{\alpha R}\rangle$ is another term in the Schmidt decomposition.
Since $\mathcal{M}$ is
defined to map
each left-hand Schmidt state to its partner in the Schmidt decomposition,
we can read off how $\mathcal{M}$ acts:
it maps $|\Phi_{\alpha L}\rangle$ to $|\Phi_{\alpha R}\rangle$ and
$\mathcal{I}|\Phi_{\alpha R}\rangle$ to $(-)^{P+\frac N2-k}\mathcal{I}
|\Phi_{\alpha L}\rangle$. Applying $(\mathcal{MI})^2$ to
$|\Phi_{\alpha R}\rangle$ therefore gives $(-1)^{\delta+k}$, for a state
with $k$ extra Fermions on the right side.

\end{document}